\documentclass[aps,prd,floatfix,superscriptaddress,preprintnumbers,
nofootinbib]{revtex4-1}
\usepackage{psfrag,graphicx,epsfig,amsmath,amssymb,bm,tikz,multirow,booktabs}
\usetikzlibrary{arrows,decorations.markings,decorations.pathmorphing}
\usepackage[font=footnotesize,justification=raggedright]{caption}

\newcommand{\eps}{\epsilon}
\newcommand{\Mkk}{M_{\mathrm{KK}}}
\newcommand{\cws}{c_w^2 }
\newcommand{\sws}{s_w^2 }

\newcommand{\ord}{\mathcal{O}}

\newcommand{\B}{\mathcal{B}}

\newcommand{\ba}{\begin{eqnarray}}
\newcommand{\ea}{\end{eqnarray}}
\newcommand{\no}{\nonumber}

\begin{document}

\preprint{FERMILAB-CONF-13-435-T}

\title{\boldmath Constraining RS Models by Future 
Flavor and Collider Measurements: \\ A Snowmass Whitepaper}

\date{\today}
\author{Kaustubh Agashe}
\affiliation{Maryland Center For Fundamental Physics, 
Department of Physics, University of Maryland, College Park, MD 20742, USA}

\author{Martin Bauer}
\affiliation{Fermi National Accelerator Laboratory, P.O. Box 500, Batavia, IL
60510, USA}
\affiliation{Enrico Fermi Institute, University of Chicago, Chicago, IL 60637,
USA}

\author{Florian Goertz}
\affiliation{Institute for Theoretical Physics,
ETH Zurich, 8093 Zurich, Switzerland}

\author{Seung J. Lee}
\affiliation{Department of Physics, Korea Advanced Institute of Science and 
Technology, \\335 Gwahak-ro, Yuseong-gu, Daejeon 305-701, Korea}
\affiliation{School of Physics, Korea Institute for Advanced Study, Seoul 
130-722, Korea}

\author{Luca Vecchi}
\affiliation{Maryland Center For Fundamental Physics, 
Department of Physics, University of Maryland, College Park, MD 20742, USA}

\author{Lian-Tao Wang}
\affiliation{Enrico Fermi Institute, University of Chicago, Chicago, IL 60637,
USA}
\affiliation{KICP and Dept. of Physics, University of Chicago, 5640 S. Ellis 
Ave., Chicago, IL 60637, USA}

\author{Felix Yu}
\affiliation{Fermi National Accelerator Laboratory, P.O. Box 500, Batavia, IL
60510, USA}


\begin{abstract}
\noindent
Randall-Sundrum models are models of quark flavor, because they
explain the hierarchies in the quark masses and mixings in terms of
${\mathcal O}(1)$ localization parameters of extra dimensional
wavefunctions. The same small numbers which generate the light quark
masses suppress contributions to flavor violating tree level
amplitudes.  In this note we update universal constraints from
electroweak precision parameters and demonstrate how future
measurements of flavor violation in ultra rare decay channels of Kaons
and B mesons will constrain the parameter space of this type of
models. We show how collider signatures are correlated with these
flavor measurements and compute projected limits for direct searches
at the 14 TeV LHC run, a 14 TeV LHC luminosity upgrade, a 33 TeV LHC
energy upgrade, and a potential 100 TeV machine.  We further discuss
the effects of a warped model of leptons in future measurements of
lepton flavor violation.
\end{abstract}
\maketitle

\section{Introduction}

While scales of quark flavor violation in generic new physics (NP)
models are already constrained to $\Lambda_\mathrm{NP}> 100 - 10^4$
TeV~\cite{Hewett:2012ns}, flavor violating processes in the bulk
Randall-Sundrum (RS) model are screened by the RS-GIM mechanism,
resulting in a relatively weak bound on the new physics scale of
$\Lambda_\mathrm{RS} \lesssim 10 $ TeV~\cite{Csaki:2008zd,
  Agashe:2004cp, Casagrande:2008hr, Bauer:2009cf}.

In the minimal realization of these models, all Standard Model (SM)
fields propagate in the five dimensional (5D) bulk, while the Higgs is
confined on or close to the infrared (IR) brane.  In this setup, there
arise large contributions to the oblique $T$ parameter, which is
sensitive to the mixing of the electroweak gauge bosons and their
heavy excitations.  Similarly, there are sizable positive corrections
to the $Zb_L\bar b_L$ coupling due to both mixing of the Z boson with
its Kaluza-Klein (KK) excitations as well as non-negligible mixing of
the fermion KK modes with the bottom quark zero
mode~\cite{Agashe:2006at}.

It is possible to capture the physics of RS models by a dual
description, in which the above effects can be understood in terms of
the mixing of elementary fields with heavy composite resonances of a
new strongly coupled nearly-conformal interaction with a confining
phase at $\Lambda_\mathrm{RS}< 10 $ TeV.  In this language, the 5D
bulk gauge group corresponds to a \emph{global} symmetry of the
composite sector and only the residual symmetry group on the UV brane
is gauged in the dual theory.  This realization allows us to
understand how the minimal bulk RS model fails to protect the $T$
parameter from large corrections, because the SM bulk gauge group does
not contain an approximate custodial $SU(2)_L\times SU(2)_R$ symmetry.
An enlarged bulk gauge group with the appropriate choice of boundary
conditions can therefore protect the $T$ parameter. In the KK
decomposed theory, this manifests itself through a cancellation
between the KK modes of the $SU(2)_L$ gauge bosons and their $SU(2)_R$
equivalents. The tension in the $Zb\bar b$ couplings can also be
resolved, if the down type quarks are demanded to transform under a
suitable $SU(2)_L\times SU(2)_R$ representation in order to suppress
both the corrections from gauge bosons, as well as from fermion
mixing.  Before the discovery of the $126$ GeV scalar and in absence
of such a symmetry, the large positive contributions to $T$ and
$g^Z_{\bar b_L b_L}$ in the minimal model could also be interpreted as
a sign of a heavy SM-like Higgs boson, but in view of the most recent
electroweak parameter fit, the $T$ parameter sets a definitive bound
of $\Lambda_\mathrm{RS}>5$ TeV.  We will therefore consider the
minimal model only for very high mass scales $\Lambda_\mathrm{RS}\in
[5,10]$ TeV, and for contrast, we will also analyze a custodially
protected bulk RS model for mass scales of $\Lambda_\mathrm{RS} \in
[2,10]$ TeV.  Strong constraints from CP violation in the $K-\bar K$
system push the compositeness scale $\Lambda_\mathrm{RS} > 8$ TeV
independent of the bounds from electroweak precision measurements. We
will consider models in which this constraint is accidentally
fulfilled by an order $1\%$ fine-tuning, but also discuss models with
a UV scale large enough to evade this bound.
\\

Section~\ref{sec:WED} will contain a short introduction to the
generation of flavor hierarchies in RS models, and introduce the
parameter space of the models discussed in this note.  In
Section~\ref{sec:TvsL}, existing constraints on the parameter space
from electroweak precision tests and CP violating observables will be
quantified and the role of tree vs. loop mediated flavor violating
processes in the RS model will be discussed. The main part of this
paper is a compilation of the reach of the next generation flavor
experiments in terms of the model parameter space, which will be given
in Section~\ref{sec:Flavor} for quark flavor observables.  In Section
\ref{sec:FvsColl} we will argue how these experiments provide a unique
possibility to constrain or discover an RS type model, because they
probe a complementary parameter space to the reach of LHC or future
collider searches.  Section~\ref{sec:Leptons} discusses the anarchic
model building considerations focusing on the leptons and
Section~\ref{sec:LeptonObservables} discusses future measurements and
constraints arising from lepton flavor observables.  We conclude in
Section~\ref{sec:Conc}.


\section{Warped Extra Dimensions}\label{sec:WED}

In order to test the RS setup
we would like to overconstrain its parameters and search for possible
inconsistencies. Here we will only introduce the most important
concepts and relations for this purpose and refer to
Refs.~\cite{Casagrande:2008hr, Bauer:2009cf, Casagrande:2010si} for a
detailed setup of the models under consideration.  We will parametrize
the extra dimension by the dimensionless variable $t\in [\epsilon,
  1]$.  The 5D metric reads
\begin{align}
   & ds^2 = \frac{\epsilon^2}{t^2}\,\left(\eta_{\mu\nu}\,dx^\mu dx^\nu 
    - \frac{1}{\Mkk^2} dt^2 \right)\,,
\end{align}
leading to a suppression of mass scales at the IR boundary at $t=1$ by
the \emph{warp factor} $\epsilon$.  This offers a solution to the
hierarchy problem, if the Higgs sector is localized at this brane and
if we chose $\epsilon$ to be given by the ratio between the Planck and
the electroweak scale,
$\epsilon=\Lambda_{\mathrm{Weak}}/\Lambda_{\mathrm{Pl}}$ .  Above,
$\Mkk= k\epsilon$ is the new physics scale, or \emph{KK scale}, and
$k$ is the curvature of the Anti-de Sitter (AdS) space. From here on,
we will identify $\Lambda_\mathrm{RS}\equiv \Mkk$. If $r$ denotes the
radius of the extra dimension, its size can also be described by its
dimensionless \emph{volume} $L=kr\,\pi\sim 36$, where $\epsilon= e^{-
  L}$.

Besides the two new universal parameters, $L$, which is fixed by the
warped geometry to address the hierarchy problem, and $M_{\rm KK}$,
the 5D Lagrangian of the minimal RS model features nine quark vector
mass parameters, given by the eigenvalues of the hermitian matrices
$\bm{M}_{Q,u,d}$, in which the subscript $Q$ denotes the mass for the
$SU(2)_L$ doublet and $u, d$ the masses of the up-type and down-type
singlets.  After KK decomposition these will determine the
localization of the zero modes along the fifth dimension and since
they are all of the order of the curvature scale we will refer to the
dimensionless ratios $c_{Q_i}= M_{Q_i}/k$, $c_{q_i}= -M_{q_i}/k$ as
the \emph{localization parameters}.  The custodial extension with an
appropriate embedding of the fermion content requires additional
matter fields and has in principle 12 quark localization
parameters. As was shown in Ref.~\cite{Casagrande:2010si}, however,
the cancellation of the corrections to the $Zb_L\bar b_L$ vertex
requires the three additional parameters to be fixed by the SM fields'
localization, so we will not treat them as free parameters.  In
addition, there are Yukawa couplings between the 5D quarks and the
Higgs, $\bm{Y_U, Y_D}$.  These cannot be simultaneously diagonalized
together with the 5D quark mass matrices and therefore they introduce
in principle 36 new free parameters.  The physical masses of the
quarks as well as the mixing angles of the CKM matrix impose nine
constraints in total on this parameter space.

Just like in the SM, hierarchical entries in the Yukawa matrices can
generate the observed structure in the physical quark masses and
mixings in the CKM matrix. A very appealing feature of the AdS metric
is however, that the overlap of the zero modes of the 5D fields with
the IR brane and therefore with the Higgs depends exponentially on the
localization parameters, and can be expressed by the so-called
\emph{profile function}
\begin{align}
 F(c)=\mathrm{sgn}(\cos \pi c)\,\sqrt{\frac{1+2c}{1 - \epsilon^{1+2c}}}\,.
\end{align}
As a result, an order one splitting between the localization
parameters can explain the hierarchies in quark masses and mixings
even for anarchic $\ord(1)$ Yukawa couplings. Following this ansatz,
the physical quark masses are given by the eigenvalues of the
effective 4D Yukawa matrices\footnote{Even though the plots in this
  work show the result of exact calculations, we will refer to the
  relevant studies 
    \cite{Casagrande:2008hr,Bauer:2009cf,Casagrande:2010si}  for the
  exact formulas and restrict the discussion here to approximations to
  leading order in $v/\Mkk$ and quark mass hierarchies.}
\begin{align} \label{Yeff}
\bm{Y}_q^\mathrm{eff}=\mathrm{diag}\left[F(c_{Q_i})\right]\bm{Y}_q\mathrm{diag}
\left[F(c_{q_i})\right]= \bm{U}_q\,\bm{\lambda}_q\,\bm{W}_q^\dagger \,,
\end{align}
in which the unitary matrices 
\begin{align}\label{U&W}
  (U_q)_{ij} \sim
   \begin{cases} 
    \frac{\displaystyle F(c_{Q_i})}{\displaystyle F(c_{Q_j})} \,; \!\!
    & i\le j \,, \\[3mm]
    \frac{\displaystyle F(c_{Q_j})}{\displaystyle F(c_{Q_i})} \,; \!\!
    & i>j \,, 
   \end{cases} ~~
   (W_q)_{ij} \sim
   \begin{cases} 
    \frac{\displaystyle F(c_{q_i})}{\displaystyle F(c_{q_j})} \,; \!\! 
    & i\le j \,, \\[3mm]
    \frac{\displaystyle F(c_{q_j})}{\displaystyle F(c_{q_i})} \,; \!\! 
    & i>j \,, 
   \end{cases}
\end{align}
rotate from the interaction to the mass eigenbasis, so that 
 $\bm{\lambda}_q$ is a diagonal matrix with 
$(\lambda_{q})_{ii}=\sqrt{2}m_{q_i}/v$. The masses are then given by 
\begin{align}\label{masses}
 m_{u_i} &  = \,  Y_\ast 
\frac{v}{\sqrt{2}}\,|F(c_{Q_i})|\,|F(c_{u_i})|\,,\,\qquad 
m_{d_i} = \,  Y_\ast \, \frac{v}{\sqrt{2}}\, 
\,|F(c_{Q_i})|\,|F(c_{d_i})|\,,
\end{align}
where $Y_\ast=f_i(\,(Y_{q})_{kl})$ is a flavor dependent function of
$\ord(1)$ matrix elements \cite{Casagrande:2008hr}. Following the
philosophy that the Yukawa couplings add no flavor structure, $Y_\ast$
should only be sensitive to the average absolute value of
$|(Y_{q})_{kl}|$, decreasing the 36 new Yukawa parameters effectively
to 2.  In the following we will moreover assume this average in the
down sector to be equal to that in the up sector, leaving us with only
one effective parameter from the Yukawa couplings.  The expressions
\eqref{U&W} and \eqref{masses} follow from a Froggatt-Nielsen analysis
\cite{Froggatt:1978nt} and are, to leading order, independent of the
KK scale $\Mkk$.  Equation \eqref{masses} provides six constraints on
the parameter space of the RS model and two further constraints can be
derived from two of the Wolfenstein parameters,
\begin{align}
 A\sim  \frac{|F(c_{Q_2})|^3}{|F(c_{Q_1})|^2|F(c_{Q_3})|}  \,, \qquad  
\lambda 
\sim \frac{|F(c_{Q_1})|}{|F(c_{Q_2})|}\,,
\end{align}
while $\rho$ and $\eta$ only depend on the Yukawa matrices. As
indicated above, in the anarchic RS model, we will only treat the
average absolute value of the Yukawa-matrix entries as a free model
parameter. Therefore, six physical masses and two Wolfenstein
parameters constrain the average size of the Yukawa couplings and the
nine quark localizations.  This leaves only one free localization
parameter, which will determine the relative global localization of
the right handed to the left handed zero modes and we can therefore
describe the anarchic (both the minimal and the custodially protected)
RS model by
\begin{align}
\Mkk,\qquad F(c_{u_3})\qquad \text{and} \qquad Y_\ast.
\end{align}
The other parameters are then fixed to {\it leading order} in $v/\Mkk$
and the quark mass hierarchies, and we are left with only 3 additional
inputs, increasing the predictivity with respect to many other
(non-anarchic) models beyond the SM.  It is remarkable that the light
quark profiles, which are fixed by their masses, also determine the
size of the couplings to KK gauge boson excitations as well as non-SM
couplings to the $Z$ and the $W^\pm$.  Therefore, flavor-changing
neutral currents (FCNCs) are not only suppressed by powers of the new
physics scale $\Mkk$, but also by the masses of the involved quarks.
This is the so-called RS-GIM mechanism and a particularly attractive
feature of RS models. Integrating out the contributions from the
exchange of all KK excitations of the photon leads for example to the
Hamiltonian
\begin{align}\label{photoncouplings}
  {\mathcal H}_\mathrm{eff}^{(\gamma)} &= \frac{2\pi\alpha}{\Mkk^2}
  \sum_{ q,q'=u,d}\,S(\bar
q_i,q_j;\bar q_k',q_l')\, Q_q Q_{q^\prime} \, \bigg\{ \frac{1}{2L} \left(
\bar
    q\gamma^\mu q \right) \left( \bar q'\gamma_\mu q'
\right)\\&\qquad\quad\quad\mbox{}
- 2\left( \bar q_L\gamma^\mu\bm{\Delta}'_Q q_L + \bar
    q_R\gamma^\mu\bm{\Delta}'_q q_R \right) \left( \bar q'\gamma_\mu
    q' \right) \notag\\
  &\quad\mbox{}\qquad\quad+ 2L \big( \bar q_L\gamma^\mu\bm{\widetilde\Delta}_Q
    q_L + \bar q_R\gamma^\mu\bm{\widetilde\Delta}_q q_R \big)
  \otimes \big( \bar q'_L\gamma_\mu\bm{\widetilde\Delta}_{Q'} q'_L +
    \bar q'_R\gamma_\mu\bm{\widetilde\Delta}_{q'} q'_R \big) \bigg\}\notag 
  \,,
\end{align}
in which $q_L$ and $q_R$ denote vectors of quark mass eigenstates in
flavor space, $\,S(\bar q_i,q_j;\bar q_k',q_l')$ is a process
dependent symmetry factor and $\Delta F=1$ FCNCs are described by the
off-diagonal elements of the matrices (for $c_{Q_i}$ and $c_{q_i}$
close to $-1/2$)
\begin{align}\label{ZMAhierarchy}
 (\Delta_Q^{'})_{ij} \sim (\Delta_Q)_{ij}  \sim \frac{1}{2}\,
F(c_{Q_i})F(c_{Q_j})\,,\notag \\
 (\Delta_q^{'})_{ij}\sim (\Delta_q)_{ij}  \sim 
\frac{1}{2}\,F(c_{q_i})F(c_{q_j})\,. 
\end{align}
The tensor structures 
\begin{align}\label{ZMAhierarchyII}
  \big({\widetilde\Delta}_{Q}\big)_{ij}  \otimes 
\big({\widetilde\Delta}_{q}\big)_{kl} \sim 
(\Delta_Q)_{ij}(\Delta_q)_{kl}\sim 
\frac{1}{4}\,F(c_{Q_i})F(c_{Q_j})F(c_{q_k})F(c_{q_l})
\end{align}
factorize only to leading order and allow for $\Delta F=2$ FCNCs.
Note that order one factors arising from the rotation matrices in
\eqref{U&W} are neglected here (see Refs.~\cite{Casagrande:2008hr, Bauer:2009cf,
  Casagrande:2010si} for details).  Using these expressions and
Eq.~\eqref{masses}, we find that the contributions of the photon KK
modes to the Wilson coefficient of the operator
\begin{align}\label{CLR}
 \mathcal{O}_{LR}=({ \bar d_L} \gamma_\mu  s_L)\, ({ \bar d_R 
\gamma^\mu}  s_R)
\end{align}
scale parametrically like
\begin{align}\label{CLRparadep}
 C_{LR}\sim \frac{\alpha}{\Mkk^2}\frac{m_dm_s}{v^2Y_\ast^2}\,.
\end{align}
This RS-GIM suppression is very effective, so that almost all flavor
observables do still allow for a new physics scale of $\Mkk\sim 1$ TeV
for most of the parameter space.  The only exception in the quark
sector is CP-violation in Kaon mixing, measured by $\eps_K$
\cite{Csaki:2008zd}.  This observable is generally large in new
physics models which allow for operators of the type in
Eq.~\eqref{CLR} which couple left- to right-handed quarks. The strong
RS-GIM suppression \eqref{CLRparadep} is partially offset in this
observable by a model-independent chiral enhancement of the matrix
elements and a sizable contribution from renormalization group running
\cite{Bagger:1997gg, Ciuchini:1998ix, Buras:2000if,Beall:1981ze,
  Gabbiani:1996hi}. Since this is the sole exception to an otherwise
excellent model of flavor, many extensions have been proposed in order
to solve this RS flavor problem \cite{Santiago:2008vq, Redi:2011zi,
  Csaki:2009wc, Bauer:2011ah}.


\section{Constraints}\label{sec:TvsL}

The parameter points considered in this study are chosen to be in
agreement with the electroweak precision parameters $S$ and $T$, the
CP violating quantities $\eps_K'/\eps_K$ and $\eps_K$, as well as with
the measurements of the $Z\bar{b}_Lb_L$ couplings. The corrections to
oblique parameters from KK gauge boson mixing lead to bounds which are
independent of the localization or representation of the fermion
fields $F(c)$ and the Yukawa couplings $Y_\ast$.  The requirement to
agree with electroweak precision tests \cite{Baak:2011ze} at
the three sigma level together with the measured Higgs mass at $126$
GeV therefore induces a definite lower bound on the KK scale of
\begin{align}
 \Mkk^\mathrm{min}\gtrsim 5\, \text{TeV}\,,\qquad   \Mkk^\mathrm{cust}\gtrsim 
2\, \text{TeV}\,,
\end{align}
in the minimal and custodial models, respectively. The CP-violating
quantities $\eps_K$ and $\eps_K'/\eps_K$ will single out parameter
points with an accidentally small phase. Corrections to the
$Z\bar{b}_L b_L$ vertex in the minimal model are directly proportional
to $1/F^2(c_{u_3})$ and the corresponding constraint will therefore
prefer parameter points with a more IR localized right-handed top
quark. 

In addition to the tree-level observables discussed here, there are
flavor-changing processes like $b\to s\gamma$, which only arise at
loop-level.  Similarly there are loop-level diagrams which can
considerably enhance the bounds from $\eps_K'/\eps_K$. Such
contributions grow with the Yukawa couplings, while the tree-level
effects are $1/Y_\ast^2$ suppressed \cite{Csaki:2008zd}. At very large
Yukawa couplings these effects can become more important than
  the tree-level constraints. Moreover, non-perturbativity can become
  an issue for large $Y_\ast$. We will therefore employ an upper
bound of $Y_\ast<3$, and thus constrain the discussion to models with
dominant leading order effects \cite{Gedalia:2009ws}.

\section{Future Measurements of Quark Flavor Observables}\label{sec:Flavor}
The next generation of flavor measurements will play an important role
in determining the possible impact of new physics and given the null
results of the first LHC run, it remains that these measurements will
show the first signs of interactions beyond the SM. We will
concentrate on FCNCs involving down-type quarks in this paper. After
the LHC shutdown, NA62 will start its first run in October 2014 and
plans to measure the branching fraction $\B(K^+\to \pi^+ \nu\bar \nu)$
with $10\%$ accuracy \cite{NA62}. The stopped Kaon experiment ORKA at
Fermilab will succeed the BNL experiments E787/949 and aims at 1000
events in this channel, improving the projected NA62 limits by $50\%$,
surpassing the current theory uncertainty. The neutral mode
$\B(K_L^0\to \pi^0 \nu\bar \nu)$ is expected to be discovered at KOTO,
the successor of the E391a experiment at KEK, by 2017, with a
long-term goal of observing 100 events at the SM rate \cite{KOTO}.
Project X, the accelerator project driving the suite of high
luminosity experiments at Fermilab, has the potential to substantially
improve both these measurements, reaching the few per-cent accuracy
level, and to measure the charged lepton channels $\B(K_L^0\to \ell^+
\ell^-)$ for the first time~\cite{Kronfeld:2013uoa}

The upgrade of the B factory in Japan, Super-KEKB, will host Belle II,
which will improve many measurements, especially for inclusive
processes such as $b\to s \ell^+\ell^-$ \cite{BELLEII}.  Future B
flavor measurements will also be performed at the LHC, where the LHCb
upgrade has the potential to follow up on the discovery of the
$B_s^0\to\mu^+\mu^-$ decay \cite{Aaij:2013aka} by the discovery of the
corresponding $B^0$ decay.

\subsection{\boldmath $K^+ \to \pi^+ \nu\overline \nu$ and $K_L^0 \to \pi^0 
\nu\overline \nu$}\label{subsec:KLKPnunu} 

The theoretically very clean neutral leptonic decays of the $K^+$ and
the $K_L^0$ are sensitive to new physics effects in $s\to d\nu\bar\nu$
transitions, which in the models considered here are only mediated by
the exchange of the $Z$ and its KK excitations\footnote{Scalar
  contributions are tiny because of the negligible neutrino masses.},
\begin{align}
   {\cal H}_{\rm eff}^{s\to d\nu\bar\nu} 
   ={  C_{\nu}^a}\,(\bar d_L\gamma^\mu s_L) \sum_l\,
    (\bar\nu_{l\hspace{0.05mm} L}\gamma_\mu\nu_{l\hspace{0.05mm} L}) 
    +{ \tilde C_{\nu}^a} \,(\bar d_R\gamma^\mu s_R) \sum_l\,
    (\bar\nu_{l\hspace{0.05mm} L}\gamma_\mu\nu_{l\hspace{0.05mm} L}) \,,
\end{align}
with Wilson coefficients
\begin{align}
\label{eq:Cnu}
    C_{\nu}^{{\rm min}} 
  &\sim \frac{\pi\alpha}{\sws\cws\,\Mkk^2} \, 
\Big[L\left(\frac{1}{2}-\frac{\sws}{3}\right)+\frac{v^2 
Y_\ast^2}{2m_Z^2}\Big]F(c_{Q_1})F(c_{Q_2}) \,,\notag\\
    \qquad
   \tilde C_{\nu}^{{\rm min}} 
   &\sim - \frac{\pi\alpha}{\sws\cws\,\Mkk^2} \Big[L\frac{\sws}{3}+\frac{v^2 
Y_\ast^2}{2m_Z^2}\Big]F(c_{d_1})F(c_{d_2}) \,,
\end{align}  
in the minimal model and
\begin{align}
\label{eq:Cnucus}
    C_{\nu}^{{\rm cust}} 
  &\sim -\frac{\pi\alpha}{\sws\cws\,\Mkk^2} \, \frac{v^2 
Y_\ast^2}{ 2m_Z^2}\,F(c_{d_1})^2 F(c_{Q_1})F(c_{Q_2}) \,,\notag\\
    \qquad
   \tilde C_{\nu}^{{\rm cust}} 
   &\sim - \frac{\pi\alpha}{\sws\cws\,\Mkk^2} \Big[L\,\cws+{ \frac{v^2 
Y_\ast^2}{2m_Z^2}}\Big]F(c_{d_1})F(c_{d_2}) \, ,
\end{align}
in the custodially protected model.\footnote{See
  Refs.~\cite{Bauer:2009cf,Casagrande:2010si,Goertz:2011nx} for a
  discussion of the changes that appear for the custodial model.}
Note that each term in the Wilson coefficients above comes with a
different phase factor, which is a function of the Yukawa couplings
and has been omitted for simplicity.  The terms proportional to
$Y_\ast$ represent contributions from the mixing of the fermions with
their KK modes, which can become important for large $Y_\ast>1$. The
mixing of the $Z$ with its excitations generates the contributions
which are not directly proportional to $Y_\ast$. Rewriting the
expressions (\ref{eq:Cnu}) and (\ref{eq:Cnucus}) in terms of
the free RS-flavor parameters $Y_\ast$ and $F(c_{u_3})$, we can see
that in the minimal model the left-handed quark currents are
generically subject to the largest corrections.  Given $Y_\ast
\lesssim 1$, these corrections scale to good approximation like
$C_\nu^{\rm min} \sim Y_\ast^{-2} F(c_{u_3})^{-2}$.  Note that only
for large $Y_\ast \sim Y_\ast^{\rm max}\sim 3$ and a strongly IR
localized top quark, $F(c_{u_3})>1$ can the corrections to the right
handed currents become dominant: these then scale like $\tilde
C_\nu^{\rm min} \sim Y_\ast^2 F(c_{u_3})^2$.


In the custodial model, fermion mixing entering the left-handed
currents is always suppressed by $m_d^2/m_Z^2$ and the contributions
of the neutral gauge bosons to these currents cancel, so that
basically only contributions to the right-handed currents remain.

\begin{figure}[t]
\includegraphics[width=0.48\textwidth]{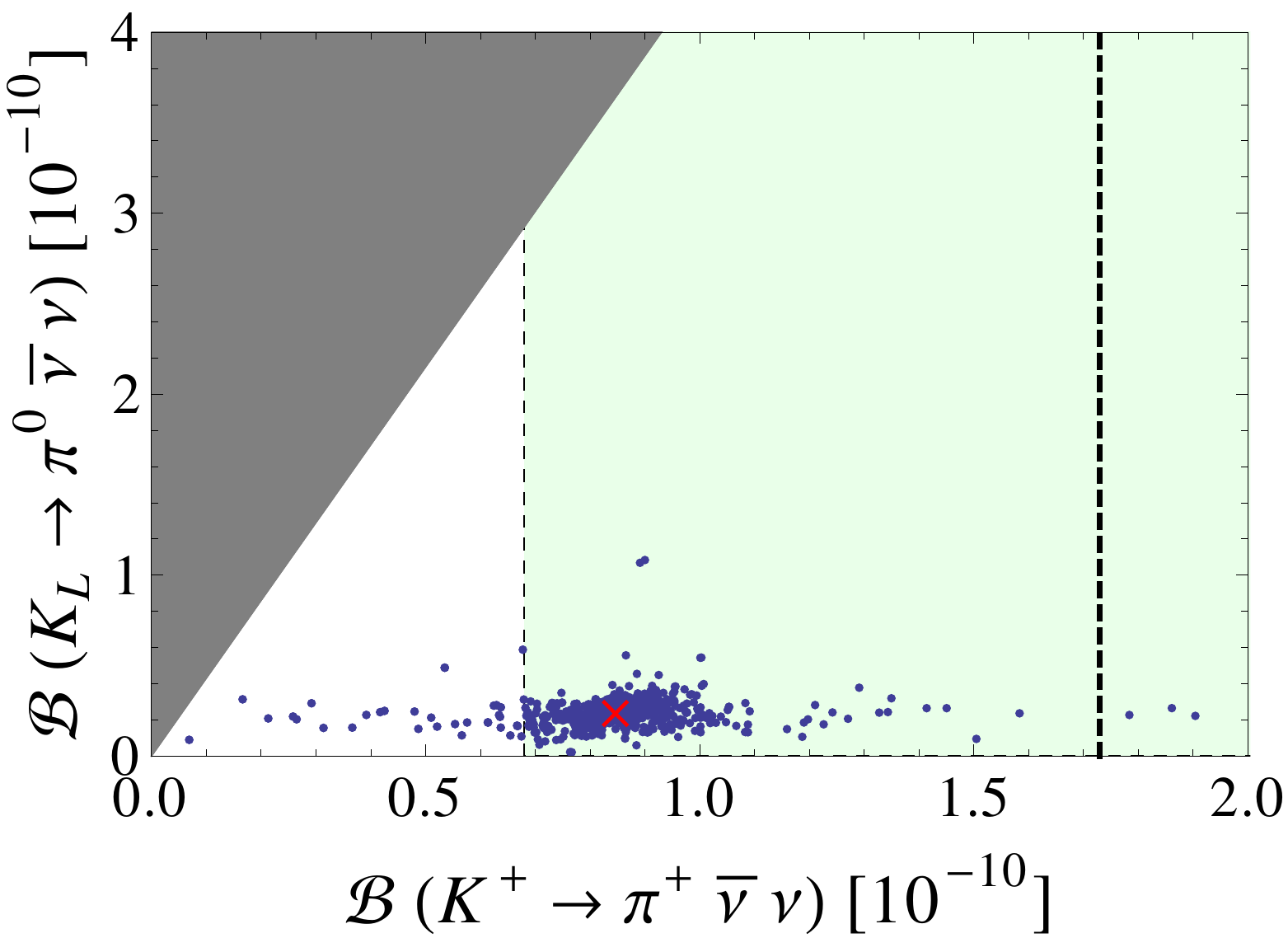}
\includegraphics[width=0.48\textwidth]{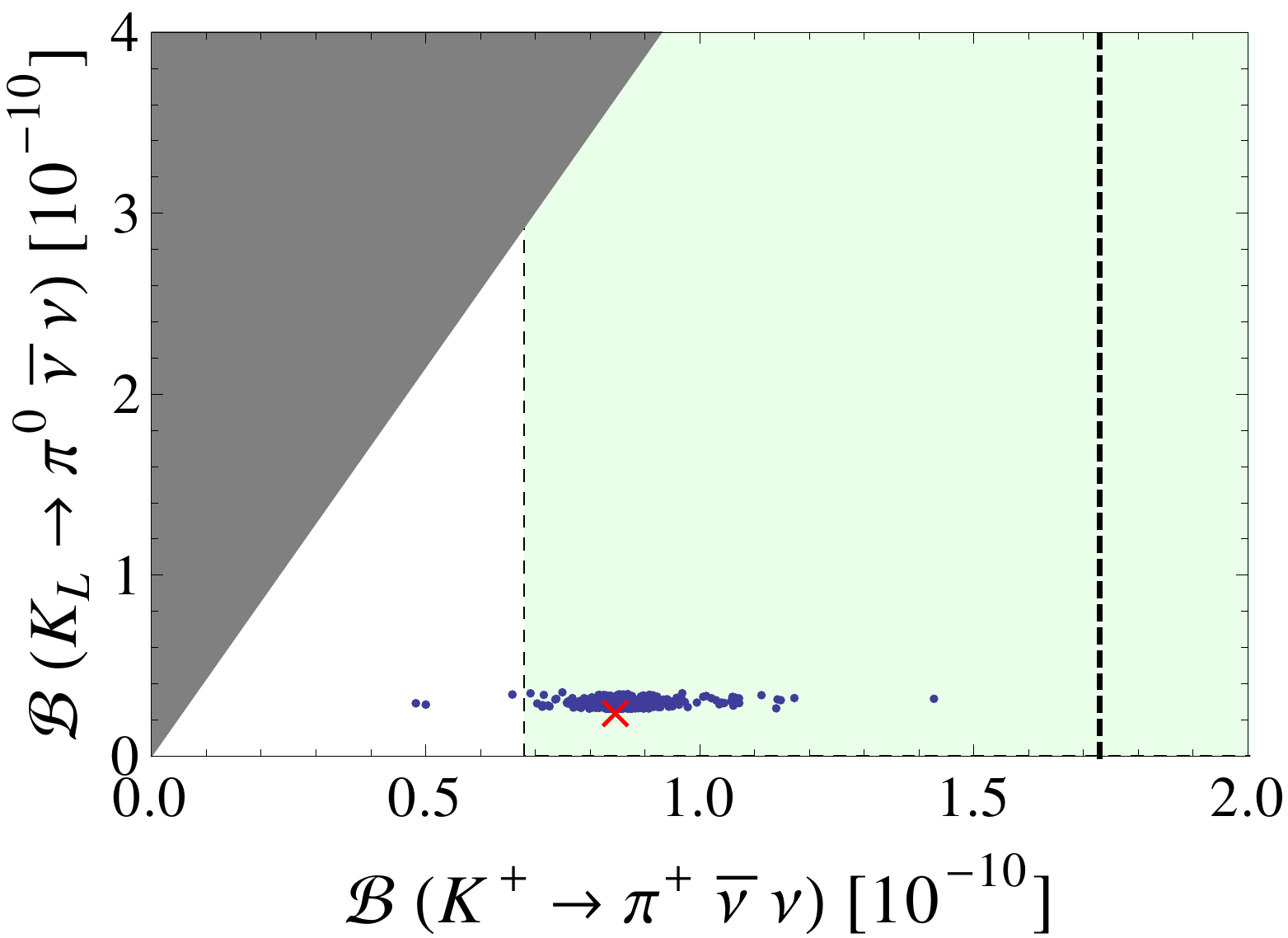} 
\caption{\raggedright Scatter points in the ${\cal B} (K_L \to \pi^0
  \nu \bar \nu)-{\cal B} (K^+ \to \pi^+ \nu \bar \nu)$ plane for the
  minimal (custodial) model on the left (right) panel. The SM
  prediction is shown as a red cross. The dark gray region is
  theoretically excluded (Grossmann-Nir bound) and the experimentally
  preferred region is shown in light blue.  }
\label{fig:BKLnunuscatterplots}
\end{figure}

We define $X\equiv 1.24\,e^{2.87\,i}+X_\mathrm{RS}^a$, in which 
\begin{align}
 X_{\rm RS}^a = \frac{ s_w^4 \cws
  m_Z^2}{\alpha^2 \lambda^5} \left (C_\nu^a + \tilde C_\nu^a \right) ,
\end{align}
and $a=\mathrm{min}$ or $a=\mathrm{cust}$. Therefore we can write
\begin{align}
   {\cal B} (K_L \to \pi^0 \nu \bar \nu) & = \kappa_L \hspace{0.5mm}
  \big ( {\rm Im} \hspace{0.5mm} X  \big )^2 \,,\label{BKLpi0nunu} \\[2mm]
  {\cal B} (K^+ \to \pi^+ \nu \bar \nu (\gamma)) & = \kappa_+
  \hspace{0.5mm} (1 + \Delta_{\rm EM} ) \hspace{0.5mm} \big | X \big
  |^2 \,,\label{BKLpi+nunu}
\end{align}
in which $\kappa_+, \kappa_L$ and $\Delta_\mathrm{EM}$ are given in
Appendix \ref{App1}.  As discussed above, the contributions from the
$Z$ and its KK modes as well as from additional neutral KK modes from
the extended gauge group lead generically to opposite chiral
structures in the minimal and custodial model.  Both observables
\eqref{BKLpi0nunu} and \eqref{BKLpi+nunu}, however, cannot resolve
this structure and therefore cannot unambiguously distinguish between
the two models. For negligible effects from fermion mixing, the
contributions in the minimal model compared to the custodial model
scale like
\begin{align}\label{XRSXCUST1}
 \frac{X_\mathrm{RS}^\mathrm{min}}{X_\mathrm{RS}^\mathrm{cust}}\sim 
\frac 1 2\,\frac{F(c_{Q_1})F(c_{Q_2})}{F(c_{d_1})F(c_{d_2})}= \lambda^{10} 
A^4\frac{m_t^4}{v^2\,m_d\,m_s}
\frac{1}{Y_\ast^2}\frac{1}{F(c_{u_3})^4}\approx \frac{6}{Y_\ast^2 
F(c_{u_3})^4}\,.
\end{align}
This leads to generically larger effects in the minimal model. The
scatter plots in Figure \ref{fig:BKLnunuscatterplots}, which scan the
parameter space in the allowed $\Mkk$ range, assuming Yukawa couplings
with $Y_\ast=1$ and $(Y_q)_{ij}<3$, illustrate this result (recall
that the mass scale for parameter points allowed by electroweak
constraints is smaller in the custodial model by roughly a factor of
two, which partially cancels this effect).

\begin{figure}[t]
\includegraphics[width=0.48\textwidth]{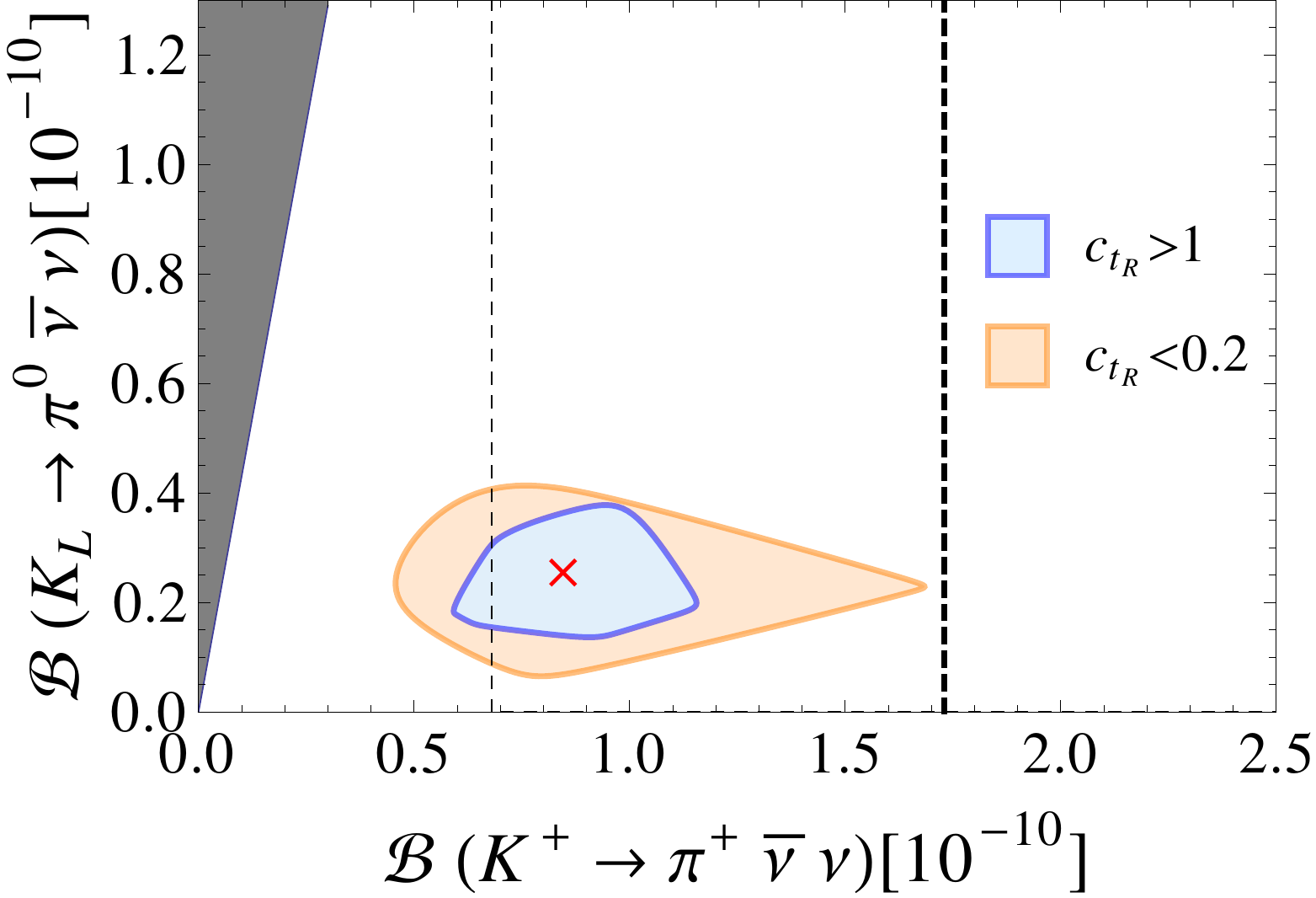}
\includegraphics[width=0.48\textwidth]{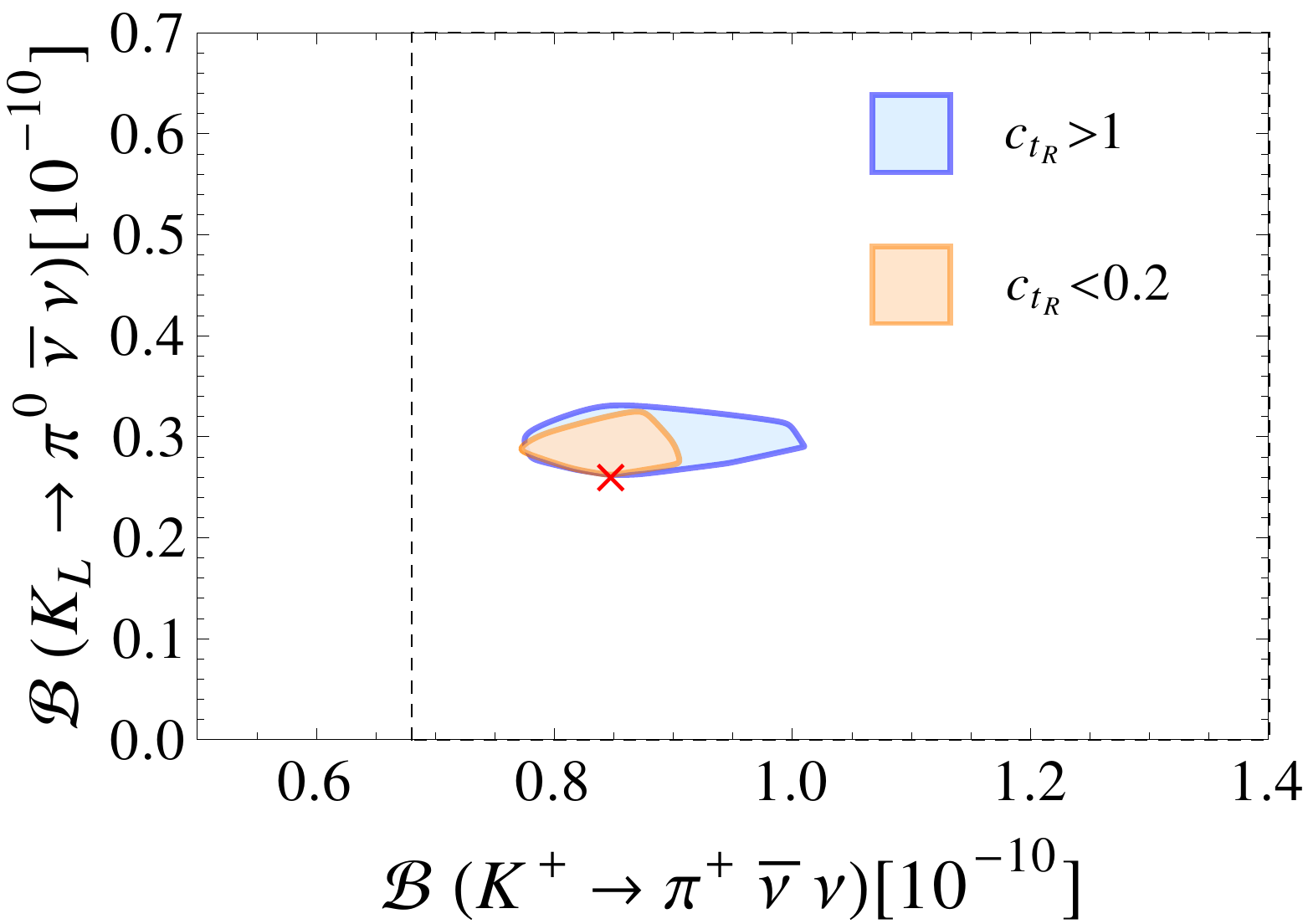} 
\caption{\raggedright Regions covering the parameter points between
  the $2.5\%$ and $97.5\%$ quantile in Fig. 1. Blue (orange ) regions
  represent $95\%$ of models with strongly (less strong) IR-localized
  right-handed tops. The SM prediction is shown as a red cross.  The
  dark gray region is theoretically excluded (Grossmann-Nir bound) and
  the central value and the one-sided one sigma limit of the
  experimental value is shown by dashed lines.  The left (right) panel
  show the size of possible effects in the minimal (custodial) model.}
\label{fig:BKLnunuplots}
\end{figure}

From \eqref{XRSXCUST1}, it is clear, that an IR shift of the
right-handed profiles into the ($c_{t_R} \equiv c_{u_3}\gtrsim 1$)
region leads to an enhanced (decreased) effect in the custodial
(minimal) model and similarly a UV shift ($c_{t_R}<0.2$) would have
the opposite effect. This behaviour is illustrated in Figure
\ref{fig:BKLnunuplots}.

The large errors on the current experimental value $\mathcal{B}(K^+
\to \pi^+ \nu\overline \nu)_\mathrm{exp}= 1.73^{+1.15}_{-1.05}\times
10^{-10}$ accommodate both models \cite{Artamonov:2008qb}.  We want to
emphasize that a future measurement with $10\%$ precision could help
to draw a much clearer picture. If new measurements point to larger
values compared to the SM, the custodially protected RS model will be
pushed to extreme corners of the parameter space\footnote{It should be
  specified, that models with different fermion representations could
  accommodate such an effect \cite{Straub:2013zca}.}.  Constraints
from $\eps_K'/\eps_K$ suppress effects in $\mathcal{B}(K_L^0 \to \pi^0
\nu\overline \nu)$ in both models (see \cite{Bauer:2009cf} for
details). A non-SM measurement of this ultra-rare decay would
therefore pose a challenge for RS type models.

%
%
%

\subsection{\boldmath  $K_L^0 \to \mu^+\mu^-$}

\begin{figure}[t]
\includegraphics[width=0.48\textwidth]{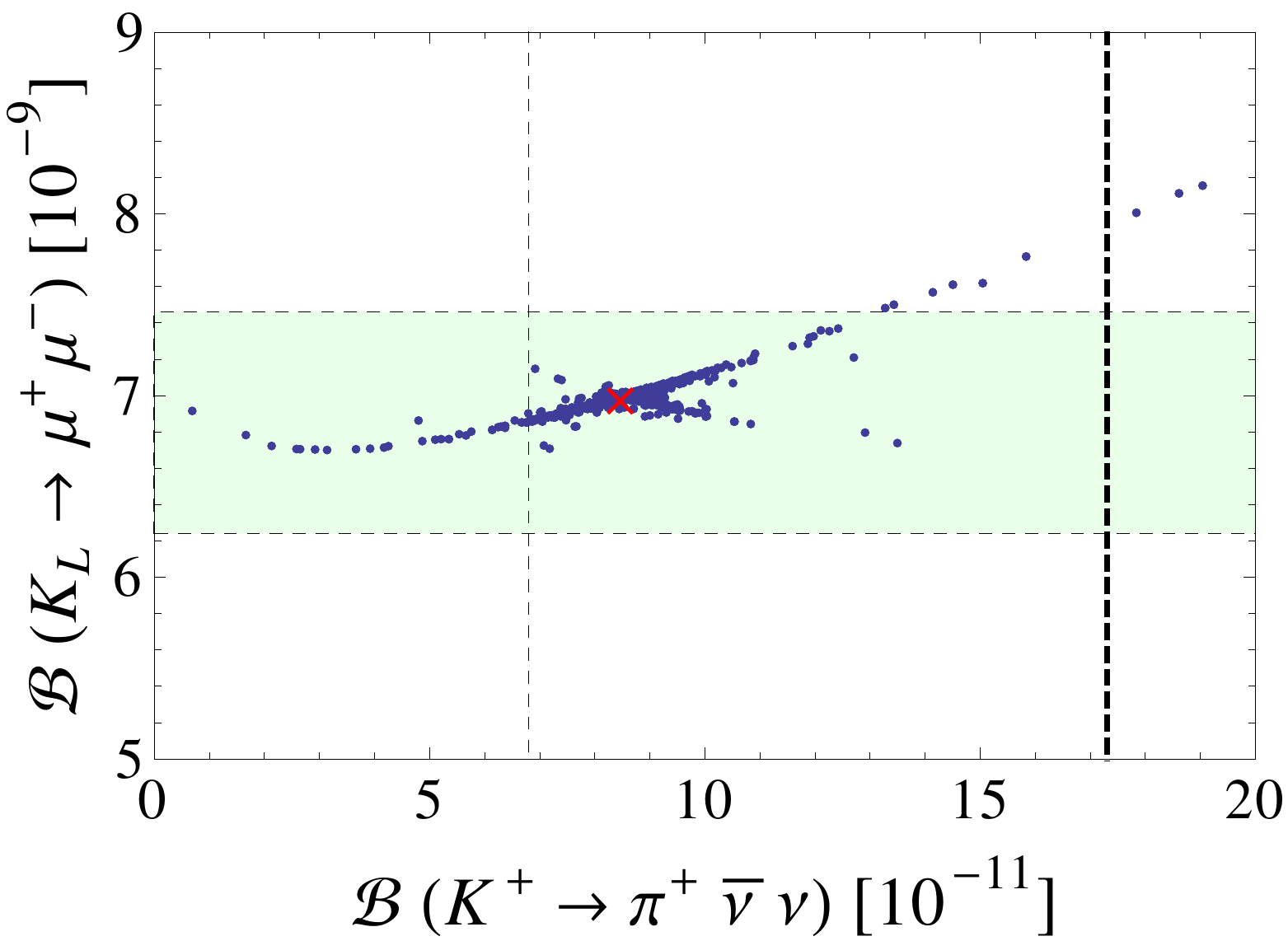} 
\includegraphics[width=0.48\textwidth]{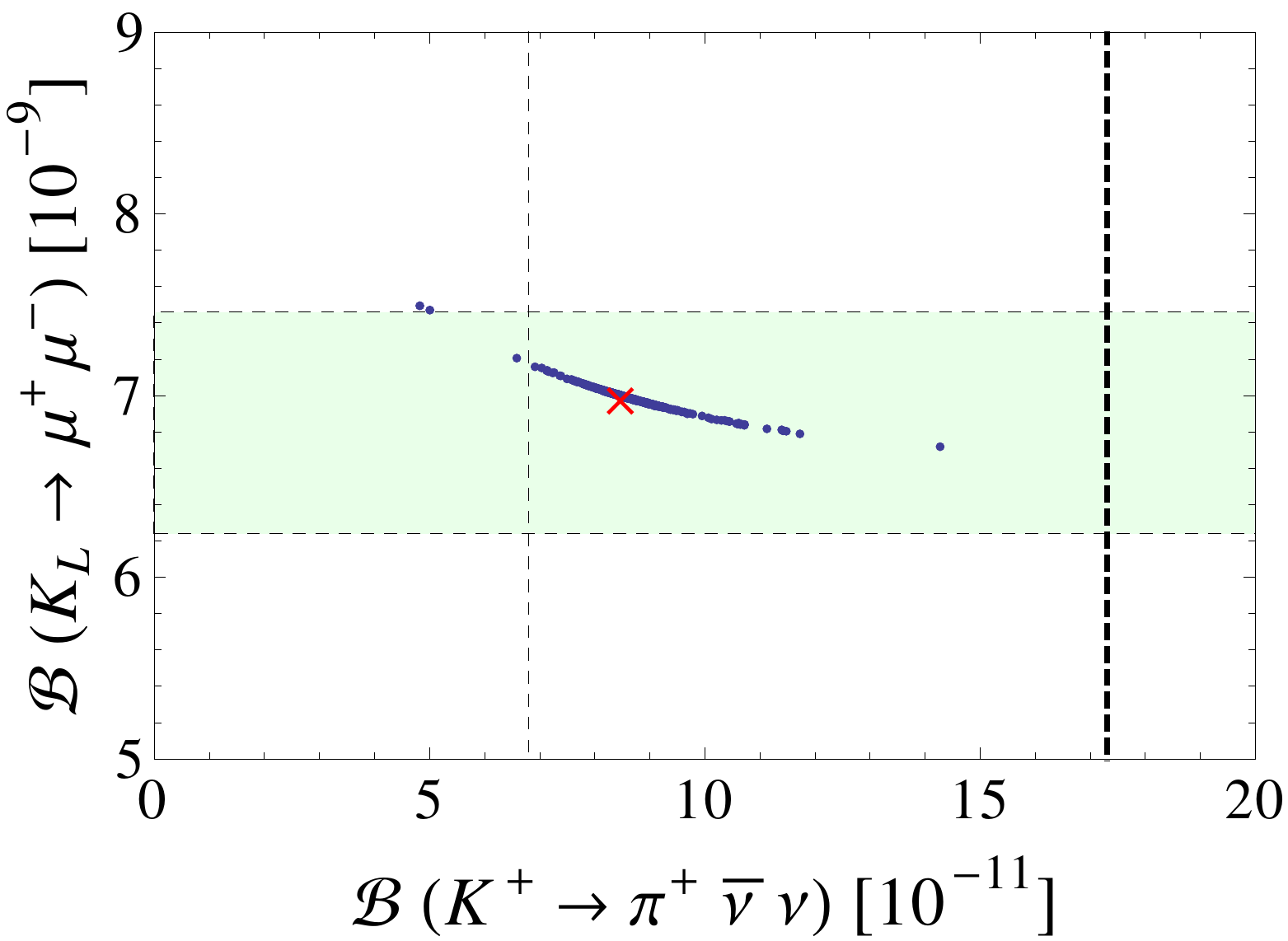} 
\caption{\raggedright Scatter points in the ${\cal B} (K_L \to \mu^+
  \mu^-)-{\cal B} (K^+ \to \pi^+ \nu \bar \nu)$ plane for the minimal
  (custodial) model on the left (right) panel. The SM prediction is
  shown as a red cross. A one sigma band combining experimental and
  theoretical errors for ${\cal B} (K_L \to \mu^+ \mu^-)$ is shown in
  light blue. The central value (and the one-sided one sigma limit)
  for ${\cal B} (K^+ \to \pi^+ \nu \bar \nu)$ is shown as a thick
  (thin) dashed line.  }
\label{fig:BKLmumuscatterplots}
\end{figure}


Contributions to decays into charged leptons are generated through
exchange of the $Z$, its excitations and photon KK modes in the
considered models. There are additional scalar contributions, which we
neglect because they are suppressed by the masses of the light
leptons.  We therefore find
\begin{align}\label{wilsonlep}
    {\cal H}_{\rm eff}^{s\to d l^+ l^-} 
   &= C_{l1}^a\,(\bar d_L\gamma^\mu s_L) \sum_l\,
    (\bar l_L\gamma_\mu l_L)
    + { C_{l2}^a}\,(\bar d_L\gamma^\mu s_L) \sum_l\,
    (\bar l_R\gamma_\mu l_R) \\
   &\quad\mbox{}+ \tilde C_{l1}^a\,(\bar d_R\gamma^\mu s_R) \sum_l\,
    (\bar l_R\gamma_\mu l_R)
    + \tilde C_{l2}^a\,(\bar d_R\gamma^\mu s_R) \sum_l\,
    (\bar l_L\gamma_\mu l_L) 
\end{align}
with the following coefficients in the minimal model,
\begin{align}\label{wilsonlepmin}
  C_{l1}^{{\rm min}} &= -\Big[\, \frac{2\pi\alpha}{3\Mkk^2}
  + 
\frac{\pi\alpha\,(1-2\sws)}{\sws\cws\,\Mkk^2}\,\left\{L\bigg(\frac{1}{2}-\frac{
\sws}{3
}\bigg)
  +\frac{v^2 Y_\ast^2}{2m_Z^2}\right\}\Big]\,F(c_{Q_1})F(c_{Q_2})
  \,, \nonumber \\
    \tilde C_{l1}^{{\rm min}} &=-\Big[  \frac{2\pi\alpha}{3\Mkk^2}\,
  +\frac{2\pi\alpha}{\cws\,\Mkk^2}\,
 \left\{ L\,\frac{\sws}{3}+\frac{v^2 
Y_\ast^2}{2m_Z^2}\right\}\Big]\,F(c_{d_1})F(c_{d_2}) \,,   
 \nonumber \\
  C_{l2}^{{\rm min}} &= -\Big[\,  \frac{2\pi\alpha}{3\Mkk^2}\,
 - 
{\frac{2\pi\alpha}{\cws\,\Mkk^2}}\,\left\{L\bigg(\frac{1}{2}-\frac{
\sws}{3
}\bigg) 
 +\frac{v^2 Y_\ast^2}{2m_Z^2}\right\}\Big]\,F(c_{Q_1})F(c_{Q_2}) \,,   
\nonumber 
 \\
  \tilde C_{l2}^{{\rm min}} &=-\Big[ \frac{2\pi\alpha}{3\Mkk^2}
  - { \frac{\pi\alpha\,(1-2\sws)}{\sws\cws\,\Mkk^2}}\,
 \left\{ L\,\frac{\sws}{3}
  +\frac{v^2 Y_\ast^2}{2m_Z^2}\right\}\Big]\,F(c_{d_1})F(c_{d_2})\,,
\end{align}
and in the case of the custodially protected model
\begin{align}\label{wilsonlepcust}
  C_{l1}^{{\rm cust}} &= -\Big[\, \frac{2\pi\alpha}{3\Mkk^2}
  -\frac{\pi\alpha\,(1-2\sws)}{\sws\cws\,\Mkk^2}\frac{v^2 
Y_\ast^2}{2 m_Z^2}F(c_{d_1})^2\Big]\,F(c_{Q_1})F(c_{Q_2})
  \,, \nonumber \\
    \tilde C_{l1}^{{\rm cust}} &=-\Big[\, \frac{2\pi\alpha}{3\Mkk^2}\,
  + \frac{2\pi\alpha}{\Mkk^2}\,
  \left\{L\,+ \frac{v^2 Y_\ast^2}{2\cws 
m_Z^2}\right\}\Big]\,F(c_{d_1})F(c_{d_2}) \,,   
 \nonumber \\
  C_{l2}^{{\rm cust}} &= -\Big[\,  \frac{2\pi\alpha}{3\Mkk^2}\,
+{ \frac{2\pi\alpha}{\Mkk^2}}\frac{v^2 
Y_\ast^2}{ 2\cws m_Z^2}F(c_{d_1})^2\Big]\,F(c_{Q_1})F(c_{Q_2}) \,,   
\nonumber  \\
  \tilde C_{l2}^{{\rm cust}} &=-\Big[\, \frac{2\pi\alpha}{3\Mkk^2}
  -   \frac{\pi\alpha\,(1-2\sws)}{\sws\,\Mkk^2}\, \left\{L\,+
\frac{v^2 Y_\ast^2}{2\cws 
m_Z^2}\right\}\Big]\,F(c_{d_1})F(c_{d_2})\,.
\end{align}
In both models the first term of each Wilson coefficient is the same
and corresponds to the contribution from the KK photon exchange. In
the custodially protected model, the contributions to the left-handed
quark-currents from the $Z$ and its excitations cancel again with the
KK excitations of the additional heavy neutral $SU(2)_R$ gauge bosons,
while the fermion mixing terms are suppressed by
$m_d^2/m_Z^2$. 

Taking into account low energy contributions, the branching ratio of
the $K_L \to \mu^+ \mu^-$ decay can be written as \cite{Mescia:2006jd}
\begin{align}
 {\cal B} (K_L \to \mu^+ \mu^-) = \left (6.7 + \big [ 1.1
  \hspace{0.5mm} {\rm Re} \hspace{0.5mm} Y_A' -0.55 \big ]^2 \right )\, \cdot 
10^{-9} \,.
\end{align}
It is only sensitive to the axial vector currents 
\begin{align}
 Y_A' = \frac{\sws\cws m_Z^2}{2\pi\alpha^2\lambda_t^{(ds)}}
    \left( C_{l1}^{a} - C_{l2}^{a} 
    + \tilde C_{l1}^{a} - \tilde C_{l2}^{a} \right)\,,
\end{align}
for $a=$ min or $a=$cust.  Therefore, the contributions from photon KK
modes cancel in both models.  Since $C_{l1}^{a} - C_{l2}^{a} = -
C_\nu^a$ and $\tilde C_{l1}^{a} - \tilde C_{l2}^{a} = \tilde C_\nu^a$,
similar to the discussion in the last subsection, in the minimal model
the left-handed currents dominate for most of the parameter space,
while the custodial protection allows only for the right-handed
currents to contribute to ${\cal B} (K_L \to \mu^+ \mu^-)$.

Figure \ref{fig:BKLmumuscatterplots} shows scatter plots in the ${\cal
  B} (K_L \to \mu^+ \mu^-)-{\cal B} (K^+ \to \pi^+ \bar\nu \nu)$ plane
for the minimal (left panel) and custodially protected model (right
panel). The expected correlations for new physics with dominant left-
(right-) handed flavor-changing currents are clearly visible in the
left (right) panel \cite{Bauer:2009cf,Straub:2013zca}.
In the light of a precise future measurement of $K^+ \to \pi^+ \bar\nu
\nu$ and similar progress in the theoretical determination of the
long-distance contributions to $K_L \to \mu^+ \mu^-$ can therefore
clearly distinguish between the models considered here. The possible
size of effects depends again on the localization of the right handed
top quark and folows a similar pattern as in Figure
\ref{fig:BKLnunuplots}. As in Section (\ref{subsec:KLKPnunu}), this
can be understood in the parametric dependence of the ratio of left-
and right-handed profiles, which control the size of effects in the
Wilson coefficients \eqref{wilsonlepmin} and
\eqref{wilsonlepcust}.\footnote{The strong correlation between the two
  observables would lead to a misrepresentation of the data in a
  two-dimensional fit based on the convex hull.  Therefore we do not
  present an illustration of this parametrical dependence.}


%
%
%

\subsection{\boldmath $B_{s,d} \to \mu^+\mu^-$}

\begin{figure}[t]
\includegraphics[width=0.48\textwidth]{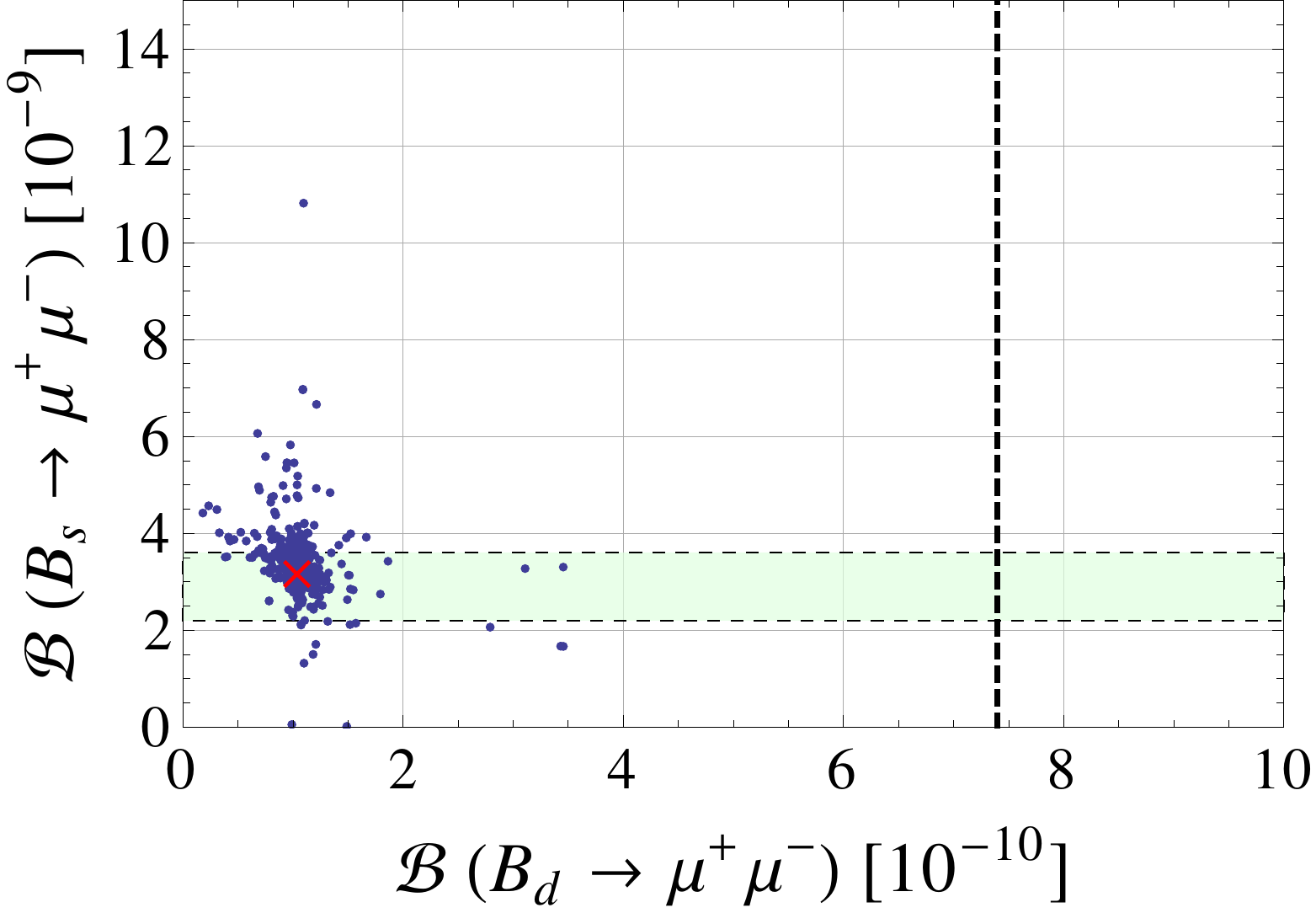}
\includegraphics[width=0.48\textwidth]{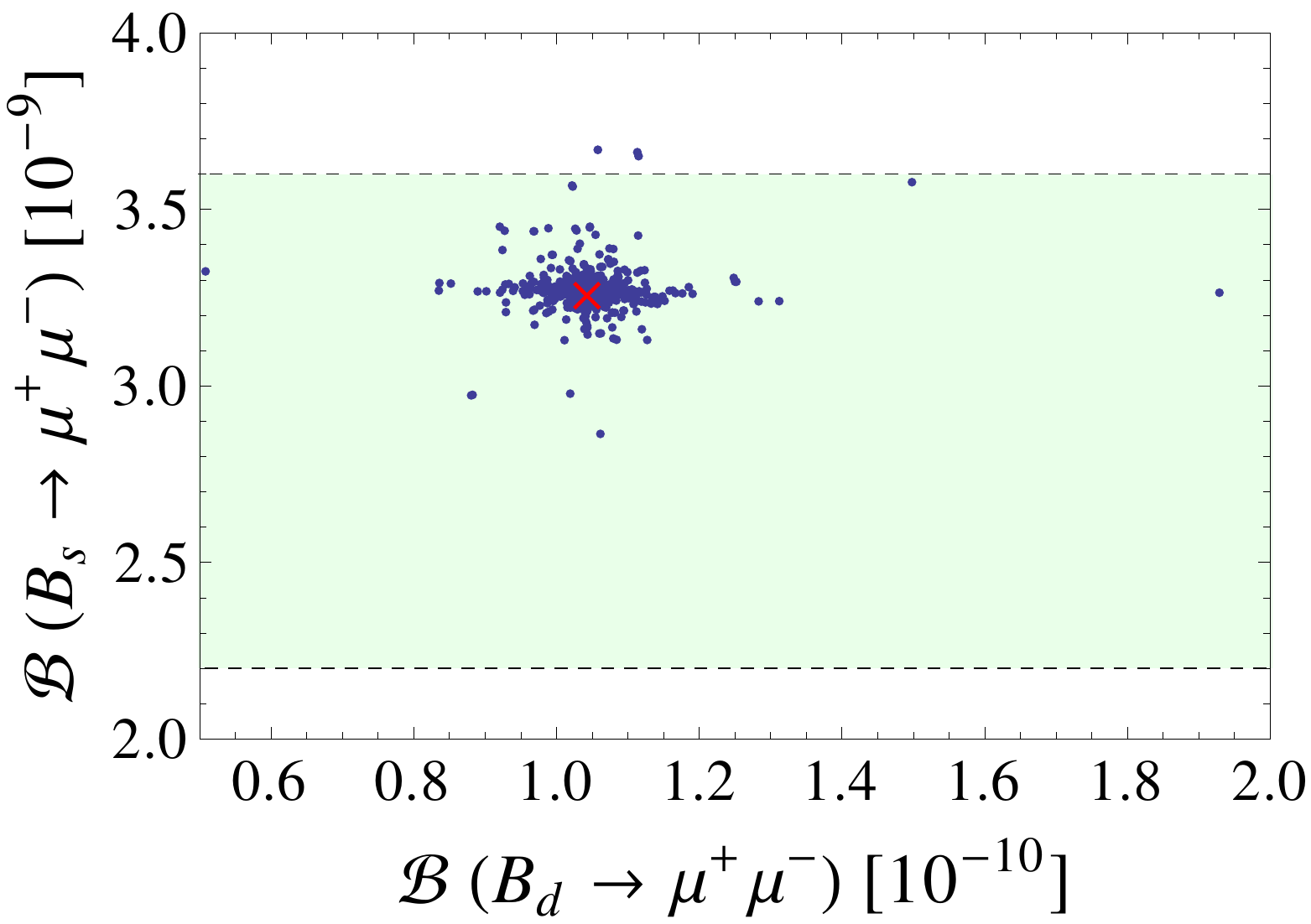} 
\caption{\raggedright Scatter points in the ${\cal B} (B_d \to \mu^+
  \mu^-)-{\cal B} (B_s \to \mu^+ \mu^-)$ plane for the minimal
  (custodial) model on the left (right) panel. The SM prediction is
  shown as a red cross. A one sigma band for the experimental value of
  ${\cal B} (B_s \to \mu^+ \mu^-)$ is shown in light blue. The upper
  limit for ${\cal B} (B_d \to \mu^+ \mu^-)$ is shown as a dashed
  line.}
\label{fig:BBmumuscatterplots}
\end{figure}

\begin{figure}[t]
\includegraphics[width=0.48\textwidth]{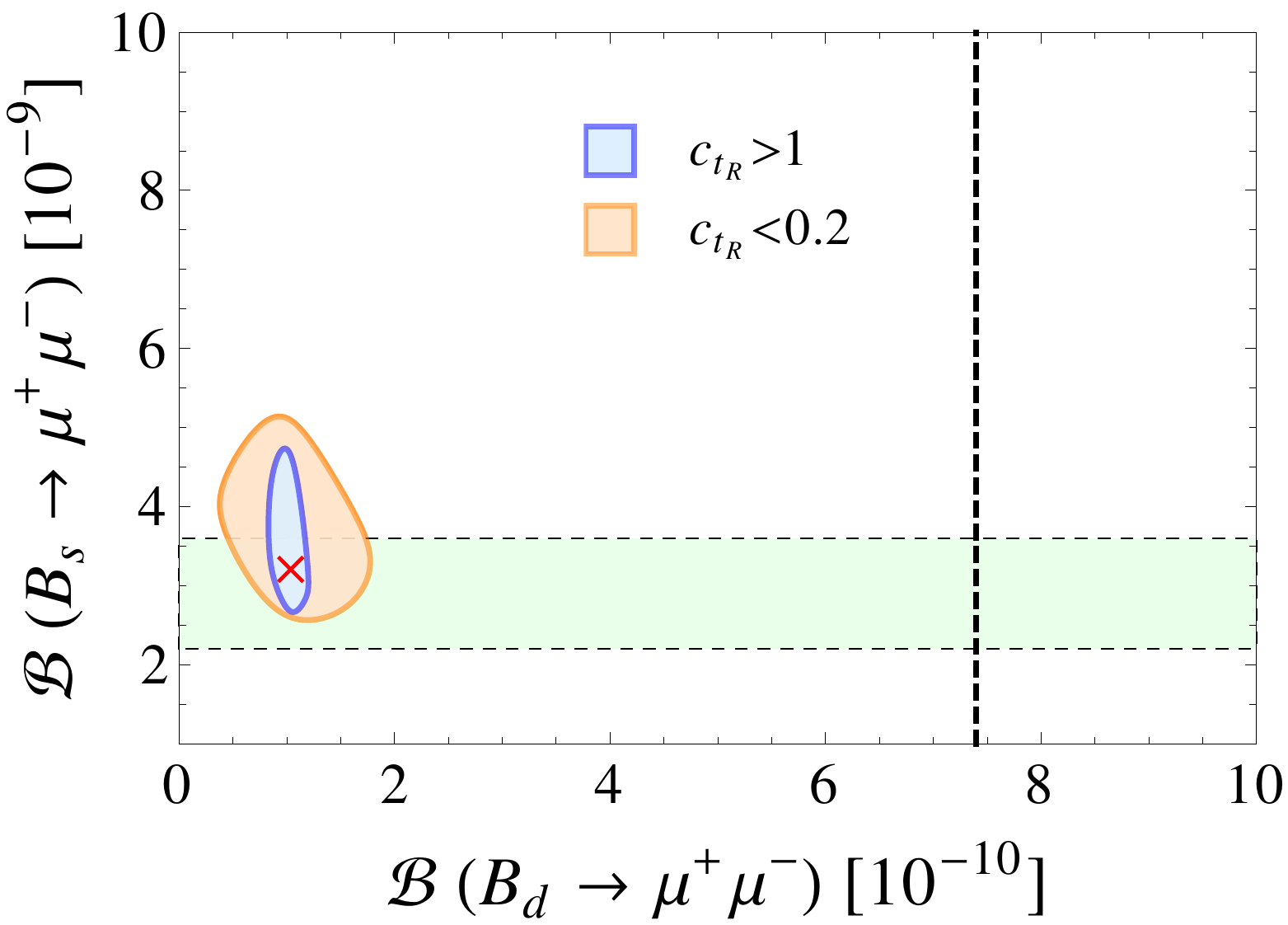}
\includegraphics[width=0.48\textwidth]{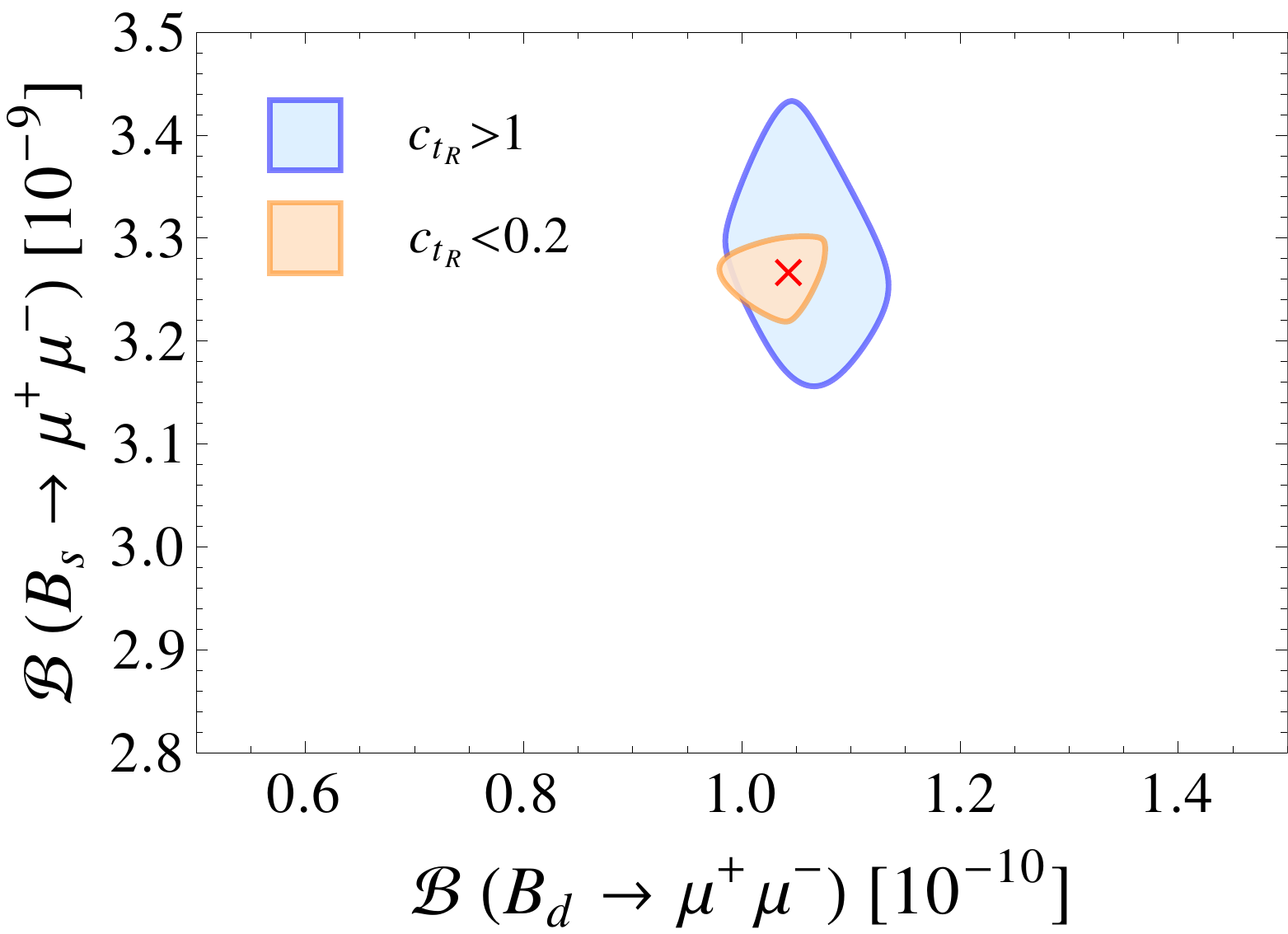} 
\caption{\raggedright Regions covering the parameter points between
  the $2.5\%$ and $97.5\%$ quantile in Fig. 5. Blue (orange) regions
  represent $95\%$ of models with strongly (less strong) IR-localized
  right-handed tops. The SM prediction is shown as a red cross The one
  sigma band of the experimental value for ${\cal B} (B_s \to \mu^+
  \mu^-)$ and the upper limit on ${\cal B} (B_d \to \mu^+ \mu^-)$ are
  shown by dashed lines.  The left (right) panel show the size of
  possible effects in the minimal (custodial) model.}
\label{BBmumuplots}
\end{figure}

The branching ratios for the $B_q \to \mu^+ \mu^-$ decays can be
expressed as

\begin{align} \label{eq:BRBqmm}
  {\cal B} (B_q \to \mu^+ \mu^-) & = \frac{G_F^2 \hspace{0.5mm}
    \alpha^2 \hspace{0.5mm} m_{B_q}^3 f_{B_q}^2 \tau_{B_q}}{64 \pi^3
    s_w^4} \, \big |\lambda_t^{(qb)} \big |^2 \, \sqrt{1 - \frac{4
      m_\mu^2}{m_{B_q}^2}}  \, \frac{4
      m_\mu^2}{m_{B_q}^2} \left | c_A + C^a \right |^2  ,
\end{align}
where $m_{B_q}$, $f_{B_q}$, and $\tau_{B_q}$ denote the mass, decay
constant, and lifetime of the $B_q$ meson (numerical values are
collected in Appendix \ref{App1}) and $c_A=0.96\pm 0.02$ is the SM
contribution \cite{Misiak:1999yg,Buchalla:1998ba}.  Again, scalar
contributions are not explicitly given, because they are additionally
suppressed by light masses. The RS contribution reads
\begin{align}
   C^a_q = - \frac{s_w^4 \cws m_Z^2}{\alpha^2\lambda_t^{(qb)}}
    \Big( C_{l1}^a -C_{l2}^a+ \tilde C_{l1}^a - 
\tilde C_{l2}^a \Big) \,.
\end{align}
The Wilson coefficients 
follow from \eqref{wilsonlepmin} for $a=$min, and from
\eqref{wilsonlepcust} for $a=$cust, with the replacements
$F(c_{Q_1})F(c_{Q_2})\rightarrow F(c_{Q_1})F(c_{Q_3})$ and
$F(c_{d_1})F(c_{d_2})\rightarrow F(c_{d_1})F(c_{d_3})$ for $B_d\to
\mu^+\mu^-$. In the case of $B_s\to \mu^+\mu^-$, we replace
$F(c_{Q_1})F(c_{Q_2})\rightarrow F(c_{Q_2})F(c_{Q_3})$ and
$F(c_{d_1})F(c_{d_2})\rightarrow F(c_{d_2})F(c_{d_3})$, as well as the
suppression of the fermion mixing terms in the custodial scenario, for
which $F(c_{d_1})^2\rightarrow F(c_{d_2})^2$.  Contributions from KK
photon exchange cancel again in both models.  In contrast to Kaon
decays, however, the ratio of left- to right-handed couplings in the
decays of $B_q$ mesons is enhanced by roughly an order of magnitude in
the minimal model, which is reflected in (again neglecting
contributions from fermion mixing)
\begin{align}
 \frac{C^\mathrm{min}_q}{C^\mathrm{cust}_q}\approx 
 \frac 1 2
\,\frac{F(c_{Q_3})F(c_{Q_i})}{F(c_{d_3})F(c_{d_i})}=\lambda^{8-2i}A^2 
\frac{m_t^4}{v^2m_b m_{d_i}}\,\frac{1}{Y_\ast^2}\frac{1}{F(c_{u_3})^4}\approx 
\frac{100}{Y_\ast^2F(c_{u_3})^4}\,,
\end{align}
with $i=1$ for $B_d$ decays and $i=2$ in the case of $B_s$ decays, and
the last expression is valid in both cases approximately. As a
consequence, effects in the custodially protected model turn out to be
even smaller compared to the minimal model, as displayed by the
scatter plots in Figure \ref{fig:BBmumuscatterplots}. The opposite
scaling behaviour with $c_{u_3}$ of effects in the minimal and
custodial model is shown by plots with fits to all parameter points in
between the $2.5\%$ and $97.5\%$ quantiles for strongly and less
strong IR localized top quarks in Figure \ref{BBmumuplots}. The
upgraded LHCb is expected to measure ${\cal B} (B_s \to \mu^+ \mu^-)$
with $8\%$ accuracy and to detect enough events in order to measure
${\cal B} (B_d \to \mu^+ \mu^-)/{\cal B} (B_s \to \mu^+ \mu^-)$ with
$35\%$ accuracy, given it is SM like. Both these measurements
will cut into the parameter space of the models considered here, even
for the high masses of the new resonances expected from electroweak
precision tests.

\section{Flavor vs Collider Observables}\label{sec:FvsColl}

\begin{figure}[t]
 \includegraphics[trim=0cm 2cm 0cm 2cm,scale=1]{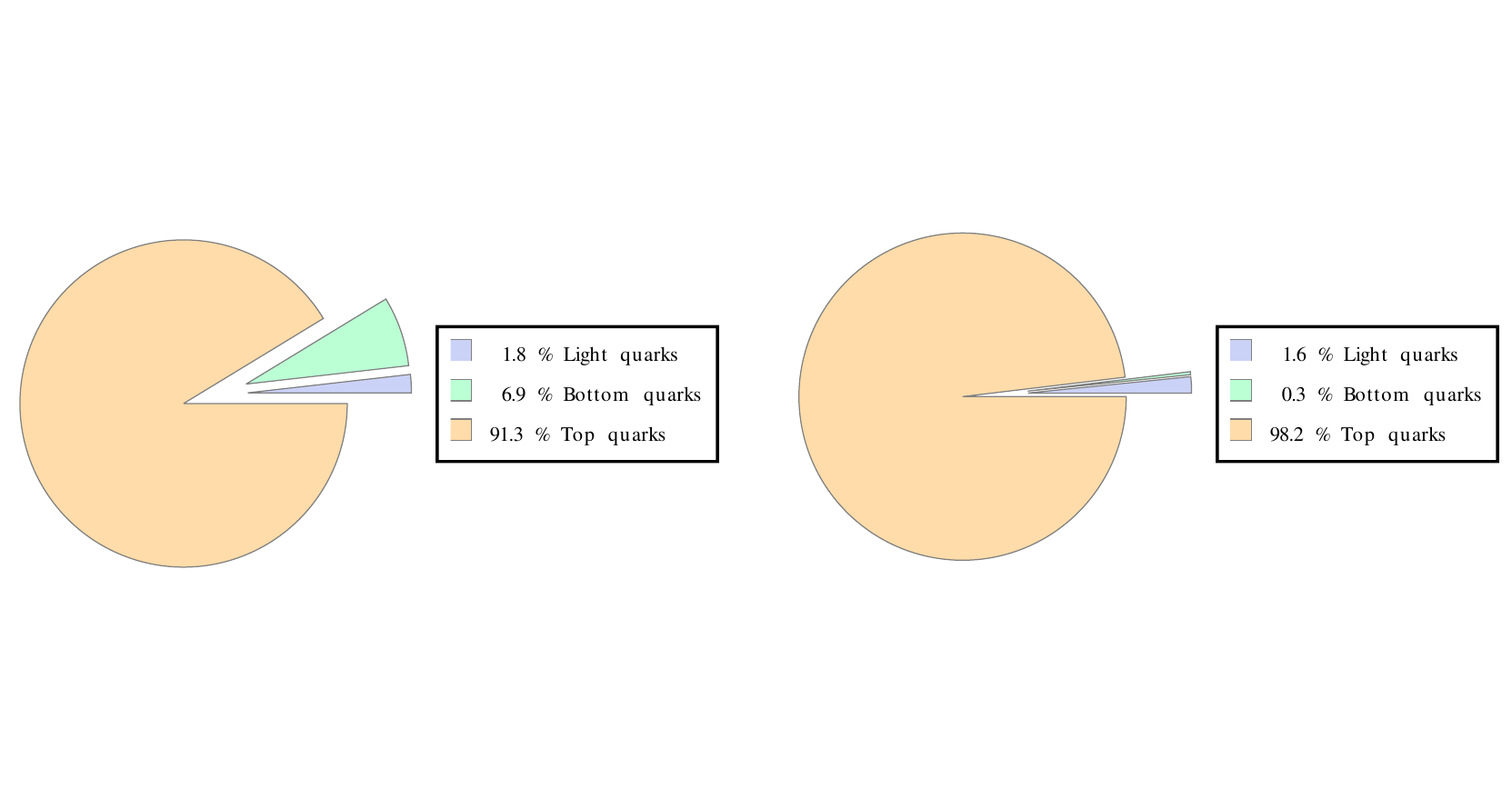}
\caption{\raggedright Pie charts of the branching fractions of the
  first KK gluon into light quarks, bottom quarks and top quarks in
  the minimal model.  The left chart shows the mean values of our
  parameter points with $c_{u_3}<0.2$, while the right chart displays
  the mean values of our parameter points with $c_{u_3}>1.5$.}
\label{BFmin.pdf}
\end{figure}

At leading order, a single KK gluon resonance cannot be produced from
gluon initial states.  The production at the LHC is therefore
suppressed by the quark-anti-quark parton density functions (PDFs) at
$\sqrt{s}=14$ TeV.  The couplings of a KK gluon to quarks is also
flavor-dependent and can be described by the same wavefunction overlap
integrals which appear in equation \eqref{photoncouplings}. The
coupling of the first gluon KK mode to quarks can be written as
\begin{align}
 g_s\,\left( g^{1ij}_{L}\, \bar{q}_i \,\gamma_\mu T^A G^\mu_A \,P_{L} 
\,q_j+g^{1ij}_{R}\, \bar{q}_i \,\gamma_\mu T^A\,G^\mu_A P_{R} \,q_j\right)\,,
\end{align}
with $i,j$ flavor indices, $G^\mu_A$ the gluon KK mode and $T^A$ the
$SU(3)$ generators. For the coupling we find 
\begin{align}\label{gkkquarks}
 g^{1ij}_{X}\approx 
\,\frac{m_{G^1}}{\sqrt{2}\Mkk}\left(\frac{1}{\sqrt{2L}}\,\delta_{ij}-\sqrt{2L}\,
(\Delta')_{ij}\right)
\end{align}
in which $m_{G^{1}}\approx 2.4\,\Mkk$ is fixed by the boundary conditions of the 
5D gluon and $\Delta' \sim F(c_{Q_i})F(c_{Q_j})$ for $X=L$ and $\Delta' \sim F(c_{q_i})F(c_{q_j})$ for $X=R$.
\footnote{Notice that this approximation of the flavor 
dependent part is 
again only valid up to order one factors and therefore only captures the 
parametric dependence of the KK gluon couplings.}
 For flavor-diagonal couplings this results in the parametric dependence
\begin{align}\label{Fsq}
 F(c_{Q_i})^2&=\frac{2m_t^2}{v^2 Y_\ast^2}\,\frac{1}{F(c_{u_3})^2}\times 
\begin{cases}\lambda^6 A^2\,,&\,\text{for}\quad i=1\\
                                                \lambda^4 
A^2\,,&\,\text{for}\quad i=2\\
                                               1\,,&\,\text{for}\quad i=3
                                               \end{cases}\,, \\
F(c_{q_i})^2&=\frac{m_{q_i}^2}{m_t^2}F(c_{u_3})^2\times 
\begin{cases}\lambda^{-6} A^{-2}\,,&\,\text{for}\quad q_i=u,d\\
                                                \lambda^{-4} 
A^{-2}\,,&\,\text{for}\quad q_i=c,s\\
                                               1\,,&\,\text{for}\quad q_i=b,t
                                               \end{cases}\,\,.
\end{align}

\begin{figure}[t]
 \includegraphics[trim=0cm 2cm 0cm 2cm,scale=1]{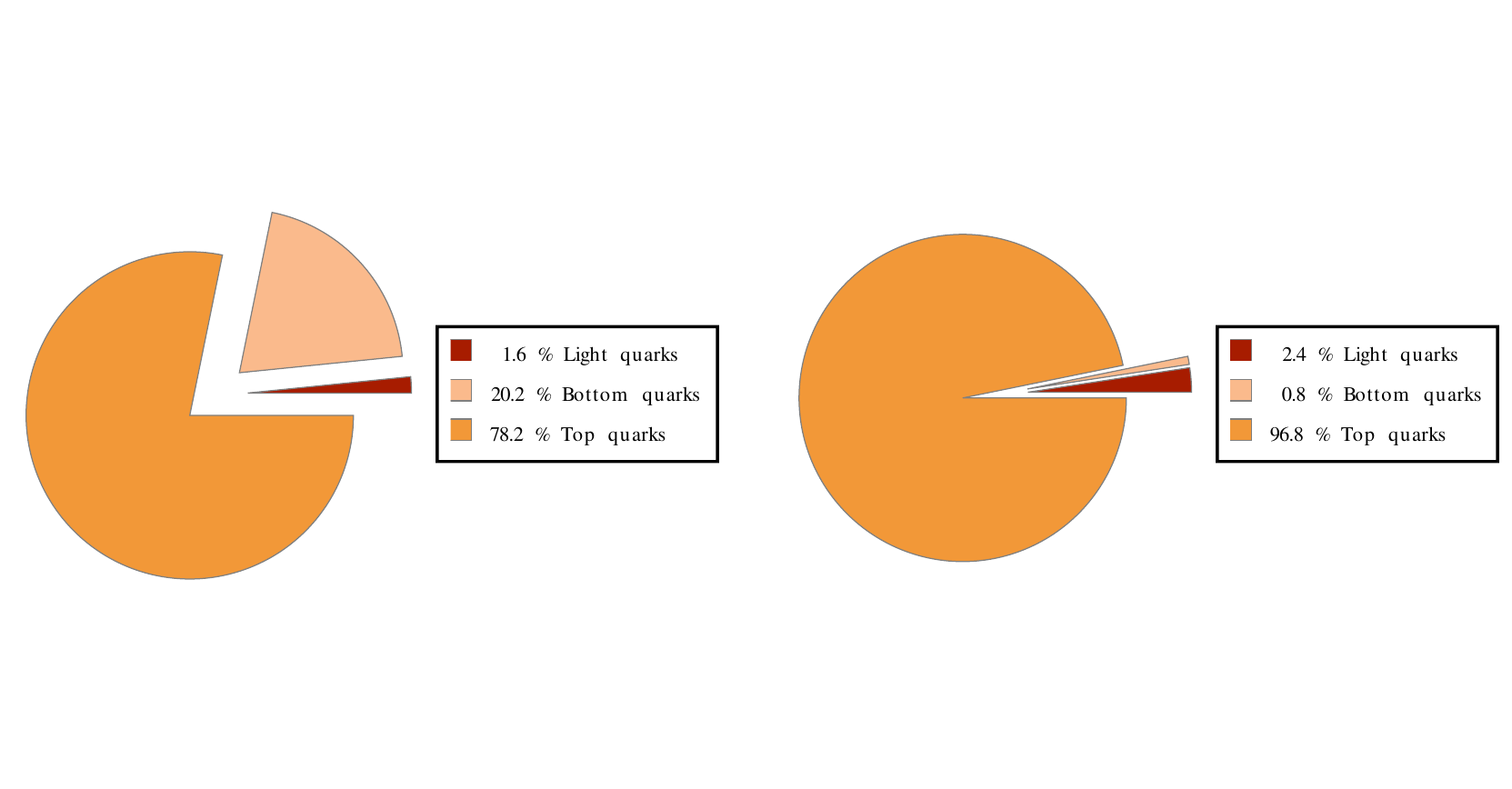}
\caption{\raggedright Pie charts of the branching fractions of the
  first KK gluon into light quarks, bottom quarks and top quarks in
  the custodial model.  The left chart shows the mean values of our
  parameter points with $c_{u_3}<0.2$, while the right chart displays
  the mean values of our parameter points with $c_{u_3}>1.5$.}
\label{BFcust.pdf}
\end{figure}

For light quarks, Eq.~\eqref{gkkquarks} is therefore clearly dominated
by the universal term, which is absent for flavor off-diagonal
couplings.  For IR localized profiles, the flavor non-universal part
dominates and can enhance the QCD coupling by a factor of a few. In
the models considered here, only the coupling to right-handed top
quarks can be subject to such an enhancement.  In the dual language,
this can be interpreted as the result of both the KK gluons and the
top quark being mostly composite and therefore strongly feel the
composite-composite couplings, whereas the elementary light quarks
couple to composites only after mixing with their resonances. As a
result, the KK gluon decays dominantly into top quarks.  From
Eq.~\eqref{Fsq} it follows that for a UV shifted right-handed top
profile, however, the coupling to the third generation electroweak
doublet can be enhanced and consequentially the partial decay width of
the KK gluon into bottom quarks can become non-negligible.
Figures~\ref{BFmin.pdf} and~\ref{BFcust.pdf} illustrate the branching
fractions of the first KK gluon excitation into light quarks, bottom
quarks and top quarks for the minimal and custodial models. The left
hand side shows the mean value of the branching fractions of our
parameter set for a UV shifted right-handed top profile
($c_{u_3}<0.2$), while the right-hand side displays the corresponding
chart for parameter points with a strongly IR localized right-handed
top ($c_{u_3}>1.5$). 

\begin{figure}[t]
\includegraphics[width=0.48\textwidth]{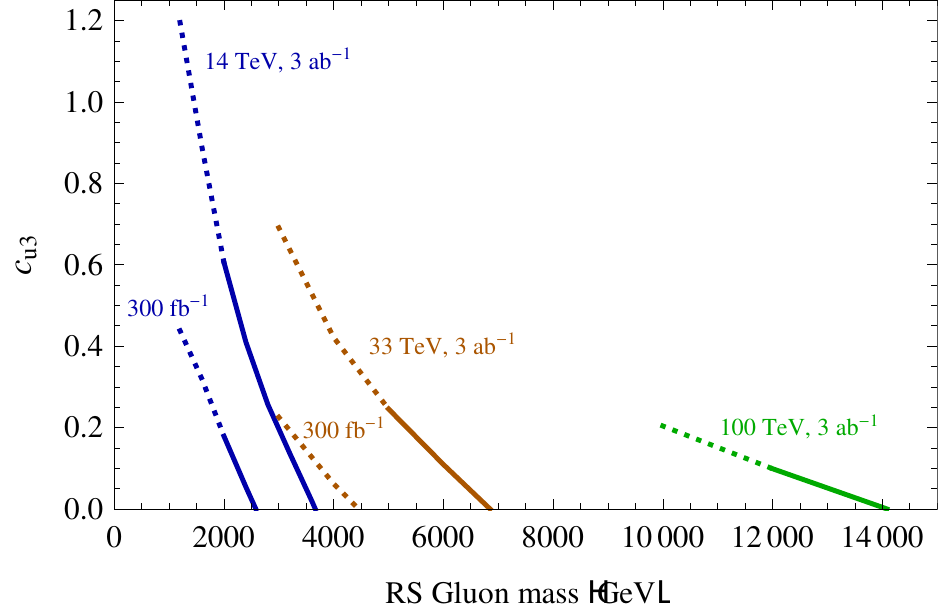}
\includegraphics[width=0.48\textwidth]{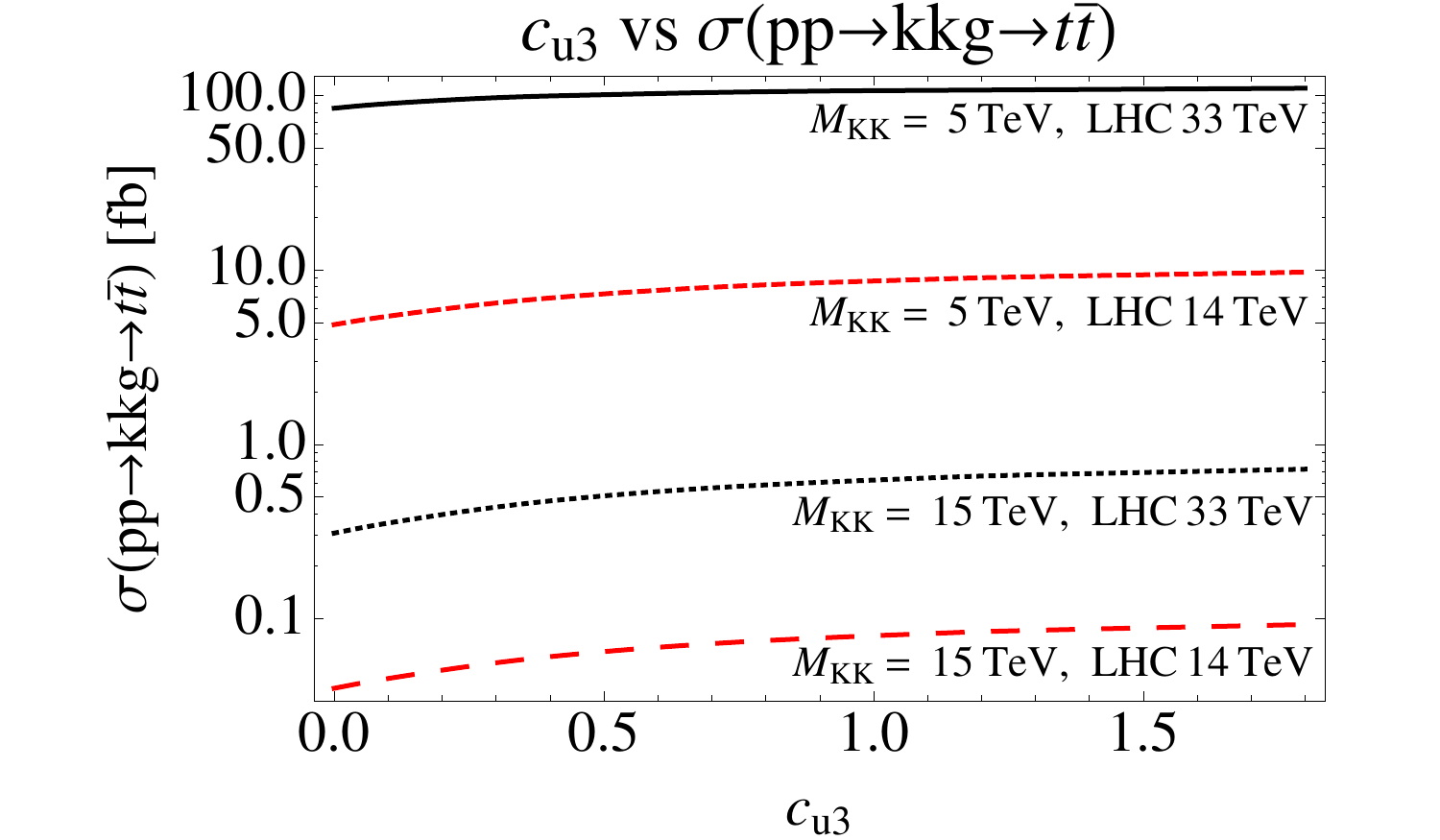}
\caption{The left panel shows projected 95\% C.L. exclusion contours
  for 14 TeV (dark blue solid), 33 TeV (dark brown solid), and 100 TeV
  (dark green solid) in the $c_{u_3}$ versus KK gluon mass plane from
  dijet resonance searches.  The regions below each line are excluded.
  The dotted lines indicate an extrapolation of the projected
  exclusion limits to low multijet trigger thresholds. The right panel
  shows the $c_{u_3}$ dependence of the cross section for the decay of
  first KK gluon excitation into $t \bar t$ pairs. }
\label{FutureKKg}
\end{figure}

The differences between the minimal and the custodial model are rooted
in the constraint from the $Z\bar{b}_L b_L$ coupling, since this
coupling increases with a more IR localized third generation
electroweak doublet and hence the constraint has a bigger impact in
the minimal model.  While a direct search for models with
$c_{u_3}>1.5$ is therefore most promising in the $t \bar{t}$ spectrum,
the $c_{u_3}<0.2$ scenario could be probed in dijet searches at an
upgraded LHC or a future collider.  In the left panel of
Figure~\ref{FutureKKg}, we show the 95\% C.L. exclusion limits
possible at 14 TeV, 33 TeV, and 100 TeV $pp$ colliders with 300
fb$^{-1}$ or 3 ab$^{-1}$ of integrated luminosity in the $M_{G}$
vs. $c_{u3}$ plane in the dijet search channel.  These limits are
based on the work of~\cite{Dobrescu:2013cmh, Yu:2013wta}, and unlike
the flavor-universal case in universal extra dimensions~\cite{Kong:2013xta}, the KK gluon mass reach
suffers because of the small production cross section and small dijet branching fraction. Interestingly,
for the custodial model the sensitivity of dijet searches grows for a
parameter region in which the sensitivity of flavor measurements
decreases.  The right panel of Figure~\ref{FutureKKg} shows that the
sensitivity in the $t \bar{t}$ final state only depends weakly on the
localization of the right handed top quark $c_{u_3}$. This is
expected, because in the considered parameter region the coupling 
of the KK gluon to top quarks grows only linearly with $c_{u_3}$.

\section{Lepton Sector: Anarchic Models}
\label{sec:Leptons}

Partial compositeness can account for the mass hierarchy in the
charged lepton sector in analogy with the quark sector.  The main
qualitative differences between the two sectors arise because: (1) the
neutrino mass matrix is non-hierarchical, and (2) flavor-changing
processes in the charged lepton sector have not been observed.

\subsection{Neutrino masses in anarchic scenarios}

We will here show how an anarchic framework can correctly reproduce
both the hierarchical charged lepton masses and the non-hierarchical
neutrino mass matrix. Other models for neutrino masses employ discrete 
symmetries
(see e.g. \cite{Csaki:2008qq, Chen:2009gy, Chen:2009hr}). 

To achieve our goal two different ways have been proposed, depending
 on whether the neutrinos are Dirac~\cite{Agashe:2008fe} or
Majorana~\cite{KerenZur:2012fr}.  Both mechanisms rely on the presence
of new, UV-sensitive sources for the neutrino mass matrix, and in turn
require new physics at or below the see-saw scale.
If neutrinos are Dirac, the RH neutrinos $N$ will also be allowed to
mix with composites of the strong dynamics. Then, up to cutoff
suppressed operators, the lepton Yukawas will be parametrized in the
IR by terms like $y_\nu\overline{\ell}HN$, with $\ell$ the lepton
doublet and \ba y_\nu\sim Y_*\left(\frac{M_{\rm KK}}{\Lambda_{\rm
    UV}}\right)^{\Delta_\ell+\Delta_{N}-5}.  \ea Here $Y_*$ denotes a
coupling between the composite states, $\Lambda_{\rm UV}$ is the UV
cutoff (assumed to be below the Planck scale), and $\Delta_{\ell,N}$
the scaling dimension of the composites mixing with $\ell,N$.  The
point raised in Ref.~\cite{Agashe:2008fe} is that if $\Delta_N$ is
sufficiently large, the above Yukawa will become so small that
cutoff-suppressed operators might become more important.  The lowest
dimension cutoff-suppressed operator we can write compatible with
lepton number involves the pair $\overline{\ell} N$, but the operator
cannot be written using the Higgs doublet because this field is a
composite state emergent in the deep IR. Yet, the strong sector
necessarily possesses an operator ${\cal O}_H$ with the same quantum
numbers of $H$, allowing us to write $\overline{\ell}{\cal
  O}_HN$. This coupling will only receive flavor-universal
renormalization effects from the strong dynamics due to the anomalous
dimension of ${\cal O}_H$, and the neutrino mass matrix will be: \ba
y_\nu\sim Y_*\left(\frac{M_{\rm KK}}{\Lambda_{\rm
    UV}}\right)^{\Delta_{H}-1}.  \ea The matrix $Y_*$ is
non-hierarchical if the physics above the cutoff $\Lambda_{\rm UV}$ is
generic, and this translates into non-hierarchical neutrino masses. In
a large $N$ theory the absence of UV sensitivity of the ``mass
operator" ${\cal O}_H^\dagger{\cal O}_H$ demands $\Delta_H\geq2$, and
to get $m_\nu=O(0.05)$ eV with $Y_*\lesssim4\pi$ we need $\Lambda_{\rm
  UV}\lesssim10^{13-14}M_{\rm KK}$.

In Ref.~\cite{KerenZur:2012fr} it has been emphasized that
cutoff-suppressed operators are always important if neutrinos are
Majorana. In this case $N$ is not there, and the leading operator
contributing to neutrino masses is $\overline{\ell^c}\tau^a\ell{\cal
  O}^a_T$, with ${\cal O}_T$ the most relevant electroweak triplet,
Lorentz scalar we can construct with the fields of the strong
sector. In the IR this results in \ba y_\nu\propto\left(\frac{M_{\rm
    KK}}{\Lambda_{\rm UV}}\right)^{\Delta_{T}-1}.  \ea and typically
requires a cutoff scale $\Lambda_{\rm UV}$ comparable to the Dirac
neutrino case. Similarly to the previous case, the neutrino masses
will be anarchic if the physics at the cutoff has no flavor structure.

There are two qualitative differences between these two
scenarios. Perhaps the most trivial one is that in this latter model
the neutrinos are Majorana, and therefore predicts neutrino-less
$\beta\beta$ decay and the standard phenomenology for Majorana
neutrinos.

At a more technical level, we find that the two models have different
TeV scale phenomenology when realized in minimal 5D scenarios. In
particular, the Majorana neutrino scenario of~\cite{KerenZur:2012fr}
requires the introduction of additional bulk scalars. Indeed, while
from the 4D point of view the existence of an electroweak triplet
composite operator is a generic expectation in strongly coupled
theories, in minimal 5D realizations a field with the required
properties is not necessarily present. A bulk Higgs pair
$H^\dagger\tau^aH$ would not do the job because its overlap with the
UV brane (where the leptons are peaked) is too small: in that case
$\Delta_{T}=2\Delta_H\gtrsim4$, and in order to generate a $m_\nu$ of
the correct order of magnitude $\Lambda_{\rm UV}$ would be required to
be unacceptably low.

The most minimal 5D realization of the scenario envisioned in
Ref.~\cite{KerenZur:2012fr} includes a bulk scalar triplet $T^a$ with
hypercharge $1$, a 4D mass $m_T\sim M_{\rm KK}$, a coupling
$\overline{\ell^c}\tau^a\ell T^a$, and a vertex $\lambda_TM_{\rm
  KK}\tilde H^\dagger \tau^aHT^a$. Once the scalar is integrated out
we get
\ba m_\nu&=&Y_T\left(\frac{m_T}{\Lambda_{\rm
    UV}}\right)^{\Delta_{T}-1}\frac{\lambda_TM_{\rm
    KK}}{m_T^2}v^2\\\no &\sim&Y_T\left(\frac{m_T}{\Lambda_{\rm
    UV}}\right)^{\Delta_{T}-2}\lambda_T\frac{v^2}{\Lambda_{\rm UV}},
\ea
where the flavor-universal factor $\sim(M_{\rm KK}/\Lambda_{\rm
  UV})^{\Delta_{T-1}}$ results from warping the coupling to leptons
down to the TeV scale. For a 5D mass close to saturating the bound
$\Delta_{T}\sim2$, we need $\Lambda_{\rm
  UV}\lesssim\lambda_T\times10^{16}$ GeV to get $m_\nu=O(0.05)$ eV.

Two more points are worth stressing. First, after electroweak symmetry
breaking, the trilinear $\lambda_T$ leads to a tadpole diagram which
results in a vacuum expectation value (vev) $v_T=\lambda_TM_{\rm
  KK}v^2/m_T^2$ for the neutral component of $T$. This in turn
contributes to the $\rho$ parameter, or equivalently: \ba \alpha_{\rm
  em} \Delta T=-4\frac{v_T^2}{v^2}=-4\left(\frac{\lambda_TM_{\rm
    KK}v}{m^2_T}\right)^2.  \ea Taking for definiteness $m_T=M_{\rm
  KK}$, we find that the bound $|\alpha_{\rm em} T|\lesssim10^{-3}$ is
easily satisfied in models where $M_{\rm KK}$ is above a few TeV even
for $\lambda_T=O(1)$. This coupling can be naturally small because it
carries custodial $SU(2)$ spurionic quantum numbers, but cannot be
arbitrarily small if we want to keep $\Lambda_{\rm UV}$ safely above
the TeV scale. The regime $m_T\ll M_{\rm KK}$ is clearly disfavored,
and suggests that the collider signatures of the triplet might well be
suppressed in a realistic model.

We also note that the presence of a vev for the triplet does not lead
to an anomalously light Majoron, the would-be Goldstone boson
associated with the breaking of the lepton number. This is because the
very same coupling $\lambda_T$ that controls spontaneous breaking of
$U(1)_L$ also controls the explicit breaking. Indeed, for
$\lambda_T=0$ the model has an unbroken (and typically anomalous)
global $U(1)_L$ and a degenerate triplet with mass $m_T$. As soon as
$\lambda_T$ is switched on, the global symmetry is explicitly broken,
and by continuity we expect the triplet masses squared to be $\sim
m_T^2$ up to a splitting of $O(\lambda_T)$.~\footnote{To rigorously
  show that the CP-odd scalar $\pi$ in $T^{(0)}=(v_T+\delta
  T)e^{i\pi/v_T}$ is not a light Majoron with ``decay constant" $v_T$
  we can employ the standard formula for the mass of a Goldstone mode
  in the presence of an anomalous current,
  $m_{\pi}^2\approx-\langle0|{\cal A}|\pi\rangle/f_\pi$. Then,
  specializing to our case, the anomaly reads ${\cal
    A}\equiv\partial_\mu J^\mu_{U(1)_L}=+i \lambda_TM_{\rm KK}\tilde
  H^\dagger \tau^aHT^a+{\rm h.c.}=-v_Tm_T^2\pi+\dots$ and the decay
  constant is $f_\pi=v_T$, so one finds $m_\pi^2\approx m_T^2$ as
  anticipated by the continuity argument.}

It is important to appreciate that the physics of the charged leptons
is essentially unaffected by the mechanism invoked to generate the
neutrinos masses. We therefore conclude that the charged lepton
observables as well as the collider phenomenology of the two
realizations~\cite{Agashe:2008fe, KerenZur:2012fr} are basically the
same.

\section{Lepton Flavor Observables in The Anarchic Scenario}
\label{sec:LeptonObservables}

We now discuss the lepton observables that are most sensitive to the
new physics scale of anarchic scenarios. The literature devoted to a
systematic study of this subject is quite modest compared to the
number of works on the quark sector, see for
example~\cite{Huber:2003tu, Agashe:2006iy, KerenZur:2012fr}.
Below we summarize the main current and projected bounds.

%
\begin{table}[h]
\centerline{\begin{tabular}{c|c|c|l}
\hline\hline
{Observable} &
  {SM Theory} &
  {Current Expt.} & 
  {Future Experiments}\\
\hline 
$a_\mu$&\multicolumn{2}{c|}{$\Delta a_\mu=(287\pm 80)\times 10^{-11}$ (E821)} &
error/4\,\, E989\\
$|d_\mu|$ &$<10^{-38}$ &$<1.9\times 10^{-19} \mathrm{e cm}
\quad\text{(E821)}$&$<10^{-21}\quad\text{ E989}$\\
${\cal B}(\mu^+ \to e^+ \gamma)$ &$\sim10^{-52}$ &$<5.7\times 10^{-13}
\quad\text{MEG 
\cite{Adam:2013mnn}}$&$<10^{-14}\quad\text{MEG\cite{Bald:2013up}}$\\
${\cal B}(\mu \to 3e)$ &$\ll 10^{-52}$ &$<1\times 10^{-12} \quad\text{SINDRUM
}$&$<10^{-16}\quad\text{Mu3e \cite{Mu3e}}$\\
$R_{\mu e}=\frac{{\cal B}(\mu A(Z,N) \to e A(Z,N))}{{\cal B}(\mu A(Z,N) \to 
\nu_\mu
A(Z-1,N))}$ &$\ll 10^{-52}$ &$<7\times 10^{-13} \quad\text{SINDRUM
II}$&$<2\times10^{-17}\quad\text{Mu2e \cite{Mu2e}}$\\
\end{tabular}}
 \caption{
  \label{tablecurrent}
  Current bounds from measurements of observables with muon initial states. 
    }
\end{table}

\begin{table}[h]
  \begin{center}
    \begin{tabular}{lllll}
      \hline \hline
Process & Current limit &\multicolumn{2}{c}{ Expected limit} & Expected limit\\
 & & \multicolumn{2}{c}{5-10 years }& 10-20 years  \\
      \hline
      $\mu^{+} \rightarrow e^{+}\gamma$ & $2.4 \times 10^{-12}$&
\multicolumn{2}{l}{$1 \times 10^{-13}$}& $1 \times 10^{-14}$ \\
      & PSI/MEG (2011) &\multicolumn{2}{l}{ PSI/MEG }& PSI,  Project X  \\
& & & \\
 \hline
      $\mu^+ \rightarrow e^+e^-e^+$ & $1 \times 10^{-12}$ & $1\times 10^{-15}$ &
$1 \times 10^{-16}$ & $1\times 10^{-17}$ \\
           & PSI/SINDRUM-I (1988) & Osaka/MuSIC & PSI/$\mu3e$ & PSI, Project X 
\\
& & & \\
\hline
$\mu^{-}N \rightarrow e^{-}N$ & $7\times 10^{-13}$ & $1\times 10^{-14} $ &
$6\times 10^{-17}$  & $1\times 10^{-18} $\\	
     & PSI/SINDRUM-II (2006) & J-PARC/DeeMee &  FNAL/Mu2e & J-PARC, Project X 
\\
     & & & \\
     \hline \hline
    \end{tabular}
      \caption{
    \label{tab:muon-lfv}
   Evolution of the 95\% CL limits on the main observables with
   initial state muons.  The expected limits in the 5-to-10 year range
   are based on running or proposed experiments at existing
   facilities.  The expected bounds in the 10-to-20 year range are
   based on sensitivity studies using muon rates available at proposed
   new facilities. The numbers quoted for $\mu^{+} \rightarrow
   e^{+}\gamma$ and $\mu^{+} \rightarrow e^{+}e^-e^+$ are limits on
   the branching fraction.  The numbers quoted for $\mu^{-}N
   \rightarrow e^{-}N$ are limits on the rate with respect to the muon
   capture process $\mu^{-}N \rightarrow \nu_\mu N^\prime$.  Below the
   numbers are the corresponding experiments or facilities and the
   year the current limit was set. Taken from
   Ref.~\cite{Hewett:2012ns}.  }
  \end{center}
\end{table}
\begin{table}[hb]
  \begin{center}
    \begin{tabular}{lll}
      \hline \hline
Process & $5\ {\rm ab}^{-1}$ & $50\ {\rm ab}^{-1}$  \\
      \hline
      ${\cal B}(\tau \to \mu\,\gamma) \rule{0pt}{2.6ex}$ &  $10 \times 10^{-9}$
&  $3 \times 10^{-9}$  \\
      ${\cal B}(\tau \to \mu\, \mu\, \mu)$ &  $3 \times 10^{-9}$ & $1 \times
10^{-9}$  \\
      ${\cal B}(\tau \to \mu \eta)$             &  $5 \times 10^{-9}$ & $2\times
10^{-9}$    \\
 \hline\hline
    \end{tabular}
  \end{center}
    \caption{
    \label{tab:LFVExptSensitivities-BelleII}
    Expected $90\%$ CL upper limits on $\tau\to\mu\gamma$, $\tau\to
    \mu\mu\mu$, and $\tau\to \mu\eta$ with $5 \ {\rm ab}^{-1}$ and $50
    \ {\rm ab}^{-1}$ data sets from Belle II and Super KEKB. Taken
    from Ref.~\cite{Hewett:2012ns}.  }
\end{table}
\begin{table}[ht]
  \begin{center}
    \begin{tabular}{lll}
      \hline \hline
    Process &  \rule{0pt}{2.6ex} Expected 90\%CL & 3$\sigma$ Evidence \\
              &  upper limit ($75$ ab$^{-1}$)   &  Reach   ($75$ ab$^{-1}$)
{\rule[-1.2ex]{0pt}{0pt}} \\
      \hline
      ${\cal B}(\tau \to \mu\,\gamma) \rule{0pt}{2.6ex}$          &  $1.8 \times
10^{-9}\
$ &  $4.1 \times 10^{-9}$  \\
      ${\cal B}(\tau \to e\,\gamma)$            &  $2.3 \times 10^{-9}$  &  $5.1
\times \
10^{-9}$  \\
      ${\cal B}(\tau \to \mu\, \mu\, \mu)$ &  $2 \times 10^{-10}$ & $8.8 \times
10^{-10}$  \\
      ${\cal B}(\tau \to e e e )$               &  $2 \times 10^{-10}$  \\
      ${\cal B}(\tau \to \mu \eta)$             &  $4 \times 10^{-10}$    \\
      ${\cal B}(\tau \to e \eta)$               &  $6 \times 10^{-10}$    \\
      ${\cal B}(\tau \to \ell K^0_S)$   {\rule[-1.2ex]{0pt}{0pt} }       &  $2
\times 10\
^{-10}$    \\
      \hline\hline
    \end{tabular}
  \end{center}
    \caption{
    \label{tab:LFVExptSensitivities}
    Expected $90\%$ CL upper limits and 3$\sigma$ discovery reach on
    $\tau\to\mu\gamma$ and $\tau\to \mu\mu\mu$ and other decays with
    75 ab$^{-1}$ at a Super$B$ type machine with a polarized electron beam. Taken
    from Ref.~\cite{Hewett:2012ns}.  }
\end{table}

\subsection{Lepton Flavor Violation in the Anarchic scenario}

We begin with $\mu\to e $ transitions and leave comments on $\tau$
physics for later.

The transition $\mu\to e\gamma$ can be described at energies smaller
than the NP scale by the following higher dimensional operators
\ba\label{dipoleL} m_\mu e
F_{\mu\nu}\left(\frac{\overline\mu_L\sigma^{\mu\nu}
  e_R}{\Lambda_L^2}+\frac{\overline\mu_R\sigma^{\mu\nu}
  e_L}{\Lambda_R^2}\right), \ea which give ($v=246$ GeV) \ba\label{BR}
{\cal B}(\mu\to e \gamma)=96\,\pi^2
e^2\left(\left|\frac{v}{\Lambda_L}\right|^4+\left|\frac{v}{\Lambda_R}
\right|^4\right).  \ea To estimate the coefficient of these operators
in anarchic models we focus on scenarios where dipoles first arise at
loop level and make use of naive dimensional analysis. Denoting by
$Y_*$ the typical coupling involved in these processes, we expect
($\langle H\rangle=v/\sqrt{2}$)
\ba
\label{Wdipole}
\frac{m_\mu}{\Lambda_L^2}\sim\frac{Y_*^2}{16\pi^2{M_{\rm
      KK}^2}}{Y_*}F(c_{\ell_2})F(c_{e_1})\frac{v}{\sqrt{2}} \ , \quad 
\frac{m_\mu}{
  \Lambda^2_R} \sim\frac{Y_*^2}{16\pi^2{M_{\rm
      KK}^2}}{Y_*}F(c_{\ell_1})F(c_{e_2})\frac{v}{\sqrt{2}}, 
\ea
where we ignore complex numbers of order unity. These naive estimates
agree with the results found in explicit 5D realizations. In those
models the relevant diagrams are generated by loops involving
intermediate fermionic KK states, and $Y_*,M_{\rm KK}$ are the 5D
Yukawa couplings and KK mass scale respectively.

Another contribution to~(\ref{dipoleL}) comes from physics at the
cutoff scale, and in a 5D setup can be parametrized by higher
dimensional operators in the bulk or localized on the IR. These in
general provide counterterms to reabsorb the divergences in the 1-loop
diagrams, and are thus expected to be of comparable magnitude by naive
dimensional analysis. In actual 5D models, the convergence of the one
loop diagrams depends on the localization of the Higgs. For example,
for a bulk Higgs the one-loop diagrams are finite and therefore
insensitive to the physics at the cutoff. In this latter case, the
cutoff effects can naturally be subleading, thus improving
predictivity.

From an inspection of~(\ref{Wdipole}), and noting that the bounds on
$\Lambda_{L,R}$ are numerically the same, follows that the optimal
scenarios are those in which
\ba\label{assume}
\frac{F(c_{\ell_i})}{F(c_{\ell_j})}\sim\frac{F(c_{e_i})}{F(c_{e_j})}
\sim\sqrt{\frac{m_i}{m_j}}.
\ea 
We will assume this relation in what follows. Incorporating the unknown $O(1)$ 
numbers in coefficients $c_{L\mu e, R\mu e}$, we define
\ba\label{Wdipole'}
\frac{m_\mu}{\Lambda_L^2}=c_{L\mu e}\frac{Y_*^2}{16\pi^2}\frac{\sqrt{m_\mu 
m_e}}{M_{\rm KK}^2} \ , \quad 
\frac{m_\mu}{\Lambda^2_R}=c_{R\mu 
e}\frac{Y_*^2}{16\pi^2}\frac{\sqrt{m_\mu m_e}}{M_{\rm KK}^2}.
\ea
The current bound from the MEG experiment~\cite{Adam:2013mnn}, also
shown in table~\ref{tablecurrent}, translates into
$\Lambda_{L,R}\gtrsim860$ TeV or 
\ba
\label{c:dipole} \frac{M_{\rm
    KK}}{Y_*}\gtrsim18\sqrt{|c_{L\mu e}|^2+|c_{R\mu e}|^2}~{\rm TeV}.
\ea
The same constraint, shown as a bound on $|c_{L\mu e,R\mu e}|$, is
presented in table~\ref{tableL}.

\begin{figure}
\begin{center}
\includegraphics[width=0.48\textwidth]{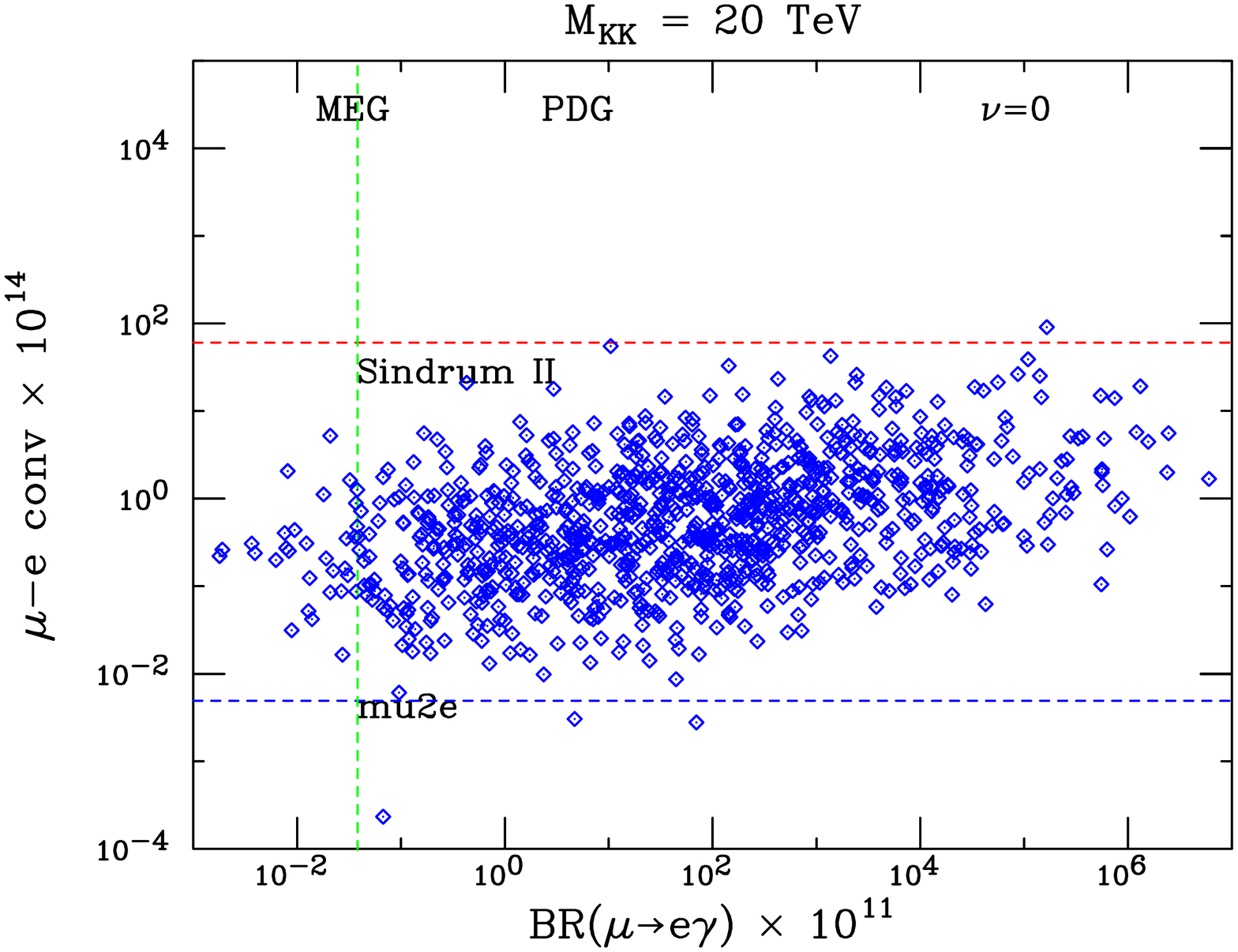}
\includegraphics[width=0.48\textwidth]{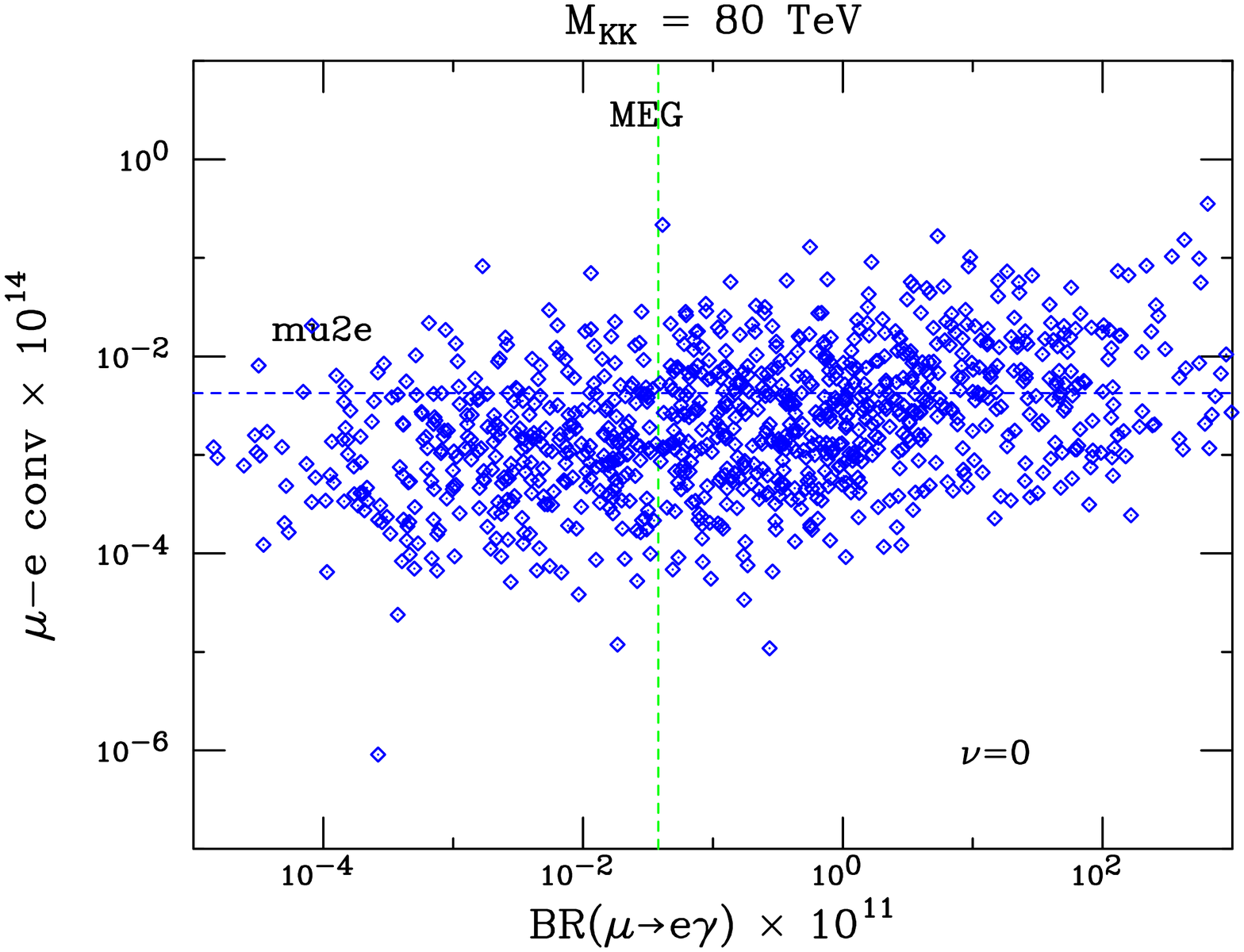}
\caption{\small Scan of the parameter space of a 5D anarchic
  model~\cite{Agashe:2006iy} with maximally spread Higgs profile
  ($\nu=0$) and satisfying Eq.~(\ref{assume}), for different
  Kaluza-Klein mass scales ($M_{KK}=20$, $80$ TeV). The dashed lines
  indicate the current $90\%$ CL bounds.
\label{scan20TeV}}
\end{center}
\end{figure}

In Fig.~\ref{scan20TeV} we compare this bound with the results of
Ref.~\cite{Agashe:2006iy} obtained for an explicit 5D model with Higgs
field with vacuum maximally spread in the bulk ($\nu=0$). The 5D
Yukawas (corresponding to our $Y_*$) are scanned in the range
$|Y_*|\in[0.5,4]$, and all points reproduce the SM lepton
masses. Fig.~\ref{scan20TeV} shows the result for $M_{\rm KK}=20$ TeV
and $80$ TeV. We find that in the upper plot only a small fraction of
points with small $Y_*$ survives the MEG bound, while for $M_{\rm
  KK}=80$ TeV a much larger parameter space becomes available. The
figure shows that the naive estimates Eq.~(\ref{Wdipole'}) agree
pretty well with the numerical results if we take
$|c_{L}|\simeq|c_R|\simeq3$ in~(\ref{Wdipole'}) (see Eq.(51)
of~\cite{Agashe:2006iy} with $\Delta_{L,R}\sim Y_* F(c_{l,e})$
and $\Delta_2\sim Y_*v$).

Another important bound on these models arises from the electron
EDM. This is generated by the CP-odd component of flavor-diagonal
operators analogous to those in Eq.~(\ref{dipoleL}), and we can
estimate
\ba \frac{d_e}{e}={\rm Im}(c_e)\frac{Y_*^2}{16\pi^2}\frac{m_e}{M_{\rm
    KK}^2}, \ea
again with $c_e = \mathcal{O}(1)$. Imposing the $90\%$ CL bound
$d_e\lesssim(6.9\pm7.4)\times10^{-28}e$ cm~\cite{Regan:2002ta} gives
$M_{\rm KK}/Y_*\gtrsim(7-36)\sqrt{|{\rm Im}(c_e)|}~{\rm TeV}$. This
bound has a much larger uncertainty compared to $\mu\to e\gamma$, but
demonstrates that both CP and flavor violation are severely
constrained in the charged lepton sector.

If CP is maximally violated the muon EDM is expected to be of order
$d_\mu\sim d_e m_\mu/m_e\sim10^{-23}e$ cm $\times(Y_*~{\rm TeV}/M_{\rm
  KK})^2$, and hence well below present and future bounds for $M_{\rm
  KK}$ in the TeV range. A similar order of magnitude estimate holds
for $(g-2)_\mu$. Electric and magnetic dipole moments of $\tau$ are
not constraining.

So far we have discussed the effect of dipole operators, and we have
seen that the constraints on the NP scale $M_{\rm KK}$ get weaker in
weakly coupled theories.  The couplings $Y_*$ cannot be taken
arbitrarily small, however, because there exist observables that get
enhanced in the limit $Y_*\ll1$. The most relevant such operator is
\ba\label{penguinL} \frac{F(c_{\ell_2})F(c_{\ell_1})}{f^2}
\overline{\mu_L} \gamma^{\mu} e_L\,i H^{\dagger}
\overleftrightarrow{D}_{\mu} H, \ea with $f\sim M_{\rm KK}/Y_*$, and
similarly for the right handed fields with $c_{\ell_i}\to
c_{e_i}$. Keeping the fermion masses $\sim Y_*F^2v$ fixed, it is clear
that a small $Y_*$ requires larger mixing parameters $F(c)$, and hence
effectively results in larger Wilson coefficients.

Let us now discuss the impact of the above operators.  Once the Higgs
acquires a vev, Eq.~(\ref{penguinL}) introduces a flavor-changing
coupling of the leptons to the $Z$ boson. Integrating out the $Z$, we
obtain 4-fermion operators with the structure
$F(c_{\ell_i})F(c_{\ell_j})\overline{\ell_i}\gamma^\mu \ell_j
J_\mu^{(Z)}/f^2$.  These are much more important than 4-fermion
operators directly generated by the strong dynamics, which are
suppressed by an additional factor of $F^2(c)$.

The induced $4$-fermion operators contribute to $\ell_i\to \ell_j$
transitions.  A stringent bound arises from $\mu\to e$ conversion in
matter. Currently, the strongest bounds come from targets of gold and
titanium~\cite{PDG}, but $R^{Au}_{\mu\to e}$ is more stringent due to
a larger coherent conversion.  Because this bound equally applies to
both chiralities of the leptons we will conservatively assume
\ba F(c_{\ell_i})\sim F(c_{e_i}).  \ea The relevant
operators can then be written as 
\ba
  \frac{c_{ij}}{Y_*f^2}\frac{\sqrt{m_im_j}}{v}~\overline{e_i}\gamma^\mu
  P_{L,R} e_j \, H^\dagger i\overleftrightarrow D_\mu H, 
\ea
 with $c_{ij}$ complex numbers of order unity. We follow
Ref.~\cite{lfv} and write
\begin{equation}
R_{\mu\to 
e}^{Au}=\left|\frac{c_{e\mu}}{Y_*f^2}\frac{\sqrt{m_em_\mu}}{v}\right|^2\left[
\left(1-4s_W^2\right) 
V_N^{(p)}-V_N^{(n)}\right]^2\frac{m_\mu^5}{\Gamma^{N}_{\rm 
capt.}},
\end{equation}
where $\Gamma^{Au}_{\rm capt.}= 8.7\times 10^{-18}$ GeV is the capture
rate in gold, while $V_{Au}^{(p)}\approx0.09$ and
$V_{Au}^{(n)}\approx0.1$ are nuclear form factors. The SINDRUM II
bound~\cite{Bellgardt:1987du} finally gives 
\ba\label{c:penguin}
f\sqrt{Y_*}\gtrsim2.0\sqrt{|c_{e\mu}|}~\rm TeV.  \ea

The observable $\mu\to ee^+e^-$ is mediated by the same
operators~(\ref{dipoleL}) and~(\ref{penguinL}), though now with the
$Z$ attached to a left and/or right $\overline{e}\gamma^\mu e$
current. The present bounds on the dipole operators are weaker than
those from $\mu\to e\gamma$, while those on the penguin
operators~(\ref{penguinL}) are comparable to those from $\mu+Au\to
e+Au$, as shown in the table.

We thus see that the most stringent constraints on the anarchic
scenarios come from $\mu\to e\gamma$ and $\mu+Au\to e+Au$, and the
next most stringent bounds are from the electron EDM and $\mu\to
e\bar{e}e$.  


Having derived the most stringent current bounds, let us turn to a
discussion of $\tau\to\ell$. The rates for these transitions can be
derived by a procedure completely analogous to the one employed
above. We therefore skip the details and merely quote the results.

For $\tau\to\ell\gamma$ we find

\ba
{\cal B}(\tau\to\ell\gamma)
&=&{\cal B}(\mu\to e\gamma)\frac{(|c_{L\tau \ell}|^2+|c_{R\tau 
\ell}|^2)}{(|c_{L\mu e}|^2+|c_{R\mu 
e}|^2)}\frac{m_\ell}{m_\tau}\frac{m_\mu}{m_e}\times {\cal B}(\tau\to 
\ell\bar{\nu}\nu).
\ea
Up to $O(1)$ numbers, the largest branching fraction is ${\cal 
B}(\tau\to\mu\gamma)\sim{\cal B}(\mu\to e\gamma)$, and is already bounded to be 
below $\sim5.7\times10^{-13}$. From this we conclude that, unless some 
cancellation among order one numbers occurs, anarchic theories satisfying 
the $\mu\to e\gamma$ bound will not show up in $\tau\to\ell\gamma$ decays.

On the other hand, the channels $\tau\to\ell\bar{f}f$ are more
promising: \ba {\cal B}(\tau\to \ell\bar{f}f)\sim{\cal B}(\mu\to
e\bar{e}e)\frac{m_\tau m_\ell}{m_\mu m_e}\times {\cal B}(\tau\to
\ell\bar{\nu}\nu), \ea where $\bar{f}f$ could be a lepton as well as a
quark pair of the same flavor. (Processes involving flavor-violating
pairs $\bar{f}f'$ are not mediated by $Z$ exchange and are suppressed
by additional powers of $F(c)\ll1$, which make them completely
negligible.)  Plugging in some representative numbers, we find ${\cal
  B}(\tau\to \mu\bar{f}f)\sim(10^2-10^3)\times{\cal B}(\mu\to
e\bar{e}e)$. Given the present bound on ${\cal B}(\mu\to e\bar{e}e)$
we argue that future B-factories will be able to probe exotic $\tau$
decays with these topologies.  The expected improvement in the
measurement of ${\cal B}(\mu\to e\bar{e}e)$ is much more dramatic,
though, and we therefore expect that among these channels $\mu\to
e\bar{e}e$ will remain the most efficient probe of anarchic
scenarios. Yet, we remark that ${\cal B}(\tau\to \mu\bar{f}f)$ will
become a complementary test of these models.

\begin{table}[ht]
\begin{center}
\begin{tabular}{c|c c|c} \hline\hline
 \rule{0pt}{1.2em}%
Operator &\quad\quad$|$Re$(c)|$ & $|$Im$(c)|$ 
& Observable \\
    \hline\hline  
  
$c\frac{Y_*^2}{16\pi^2}\frac{m_e}{M_{\rm 
KK}^2}~\overline{e}\sigma^{\mu\nu}eF_{\mu\nu}
e_{L,R}$    & --- & {$1.1~\left(\frac{m_\rho/Y}{10~{\rm TeV}}\right)^2$} &
  electron EDM
  \\
  \hline

$c\frac{Y_*^2}{16\pi^2}\frac{\sqrt{m_em_\mu}}{M_{\rm 
KK}^2}~\overline{\mu}\sigma^{
\mu\nu}eF_{\mu\nu}e_{L,R}$    &  
\multicolumn{2}{c|}{$0.31~\left(\frac{m_\rho/Y}{10~{\rm TeV}}\right)^2$} &
  $\mu\to e\gamma$
  \\
   \hline
   
    "    &  \multicolumn{2}{c|}{$3.7~\left(\frac{m_\rho/Y}{10~{\rm 
TeV}}\right)^2$} &
  $\mu\to ee^+e^-$
  \\
   \hline
   
  $\frac{c}{Y_*f^2}\frac{\sqrt{m_em_\mu}}{v}~\bar e\gamma^\mu \mu_{L,R}\, 
H^\dagger i\overleftrightarrow D_\mu H$    &   
\multicolumn{2}{c|}{$0.25~\left(\frac{\sqrt{Y}f}{1~{\rm TeV}}\right)^2$}  &
 $\mu(Au)\to e(Au)$ 
 \\ 
 
  \hline
   
  "    &   \multicolumn{2}{c|}{$0.38~\left(\frac{\sqrt{Y}f}{1~{\rm 
TeV}}\right)^2$}  &
 $\mu\to ee^+e^-$
 \\  

   \hline\hline
\end{tabular}

\vspace{0.3cm}

\caption{\label{table:generic}\footnotesize{Upper bounds on the
    dimensionless coefficients $c$ of the operators in the first
    column, assuming that $m_\rho,Y$ are real. See the text for
    details.  } }\label{tableL}
\end{center}
\end{table}

\section{Conclusions}\label{sec:Conc}
In the first part of this note, we computed the impact of resonances
in a complete model of warped extra dimensions on flavor changing
observables in the quark sector. Assuming anarchic order one Yukawa
couplings, these models can be described by one localization parameter
and the new physics scale $\Mkk$.  Universal bounds from
electroweak precision data and the measurement of the Higgs mass 
 constrain this mass scale to $\Mkk\gtrsim 5$ TeV in a minimal model
with only a SM bulk gauge group and  $\Mkk\gtrsim 2$ TeV if a
custodial $SU(2)_R\times SU(2)_L$ bulk symmetry protects the $T$
parameter from large corrections.  Since the lightest KK gluon mode
has a mass of $m_{\mathrm{G}^{(1)}}\approx 2.5\, \Mkk$ and has
small couplings to light quarks, it is challenging to observe this
state in direct collider searches.  We emphasized that precise
measurements of flavor violation in $K\rightarrow \pi \bar\nu \nu$ as
proposed by the ORKA collaboration together with future LHCb
measurements have the potential to determine the quark localization
parameters.  A future precise measurement of the correlated quantities
${\cal B}(K_L\rightarrow \mu^+\mu^-)$ and ${\cal
    B}(K^+\rightarrow \pi^+ \bar\nu \nu )$ can identify the helicity
structure of new physics contributions and therefore allows to
identify the underlying bulk gauge group.  New limits on these
quantities serve as a signpost for future collider searches for KK
resoances.  We derived the correlation between flavor and collider
observables and translated them into projected bounds at the $14$ TeV
LHC, a luminosity upgrade of the 14 TeV LHC, as well as a potential
$33$ TeV upgrade and a future $100$ TeV $pp$ machine.

In the second part of this note, we discussed the effects in charged
lepton flavor observables in models with warped extra dimensions.  We
calculated the parametric dependence of the new physics contribution
to electric dipole moments, $\mu \rightarrow e e^+e^-$, $\mu
\rightarrow e \gamma$, and $\mu \rightarrow e$ on the Yukawa
couplings, localization parameters and $\Mkk$.  The latter two
observables, $\mu \rightarrow e \gamma$ and $\mu \rightarrow
e$, provide the most stringent bounds and are already sensitive
to very high KK scales of $\Mkk >20$ TeV for anarchic order one Yukawa
matrices. The proposed next generation of experiments from the Mu2e
and MEG collaborations will probe warped models of flavor in the
lepton sector up to $\Mkk\gtrsim 80$ TeV in the anarchic case and
consequentially determine possible flavor structures in the lepton
sector if the KK scale is much lower. Finally we discussed corrections
to $\tau$ decays in Randall-Sundrum models and concluded that the
current as well as expected future bounds are generically less
constraining than the light lepton flavor observables.

\section{Acknowledgements}

Fermilab is operated by Fermi Research Alliance, LLC under Contract
No.  De-AC02-07CH11359 with the United States Department of
Energy. Some of the authors (MB, FG, LW and FY) want to thank KITP
Santa Barbara for warm hospitality and support during completion of
this work. This research is supported in part by the National Science
Foundation under Grant No. NSF PHY11-25915. KA is supported by the National 
Science Foundation
under NSF Grant No.~PHY-0968854. MB acknowledges the
support of the Alexander von Humboldt Foundation. FG acknowledges
support by the Swiss National Foundation under contract SNF
200021-143781. SL is supported by the 
National Research Foundation of Korea(NRF) grant funded by the Korea 
government(MEST) N01120547.
LV was supported in part by NSF Grant No. PHY-0968854, by the NSF Grant No. 
PHY-0910467,
and by the Maryland Center for Fundamental Physics.

\appendix

\section{Numerical Input}\label{App1}
\begin{table}[ht]
 
\begin{tabular}{c|c|c}
Paramter & Value & Reference\\\midrule[.7pt]
 $\kappa_L $&$ (2.231 \pm 0.013) \cdot 10^{-10}\hspace{0.5mm}
(\lambda/0.225)^8$ & \cite{Mescia:2007kn}\\
  $\kappa_+ $&$ (0.5173 \pm 0.0025) \cdot 10^{-10}
\hspace{0.5mm} (\lambda/0.225)^8$ & \cite{Mescia:2007kn}\\
$\Delta_{\rm EM} $&$ -0.003$ & \cite{Mescia:2007kn}\\\hline
$m_{B_d}$&$5.2796\,$ GeV &\cite{Beringer:1900zz}\\
$\tau_{B_d}$&$(1.519\pm0.007)\,$ ps & \cite{Beringer:1900zz}\\
$f_{B_d}$&$(190.5\pm 4.2)\,$ MeV&\cite{FLAG}\\
$m_{B_s}$&$5.3667\,$ GeV &\cite{Beringer:1900zz}\\
$\tau_{B_s}$&$(1.516\pm0.011)\,$ ps & \cite{Beringer:1900zz}\\
$f_{B_s}$&$(227.7\pm 4.5)\,$ MeV&\cite{FLAG}
\end{tabular}
\caption{Parameters used in the calculations in Section \ref{sec:Flavor}.}
\end{table}

\clearpage


\end{document}